\documentclass[lettersize,journal]{IEEEtran}
\usepackage{amsmath,amsfonts}
\usepackage{algorithmic}
\usepackage{algorithm}
\usepackage{array}
\usepackage[caption=false,font=normalsize,labelfont=sf,textfont=sf]{subfig}
\usepackage{textcomp}
\usepackage{stfloats}
\usepackage{xurl}
\usepackage{verbatim}
\usepackage{graphicx}
\usepackage{cite}
\usepackage{booktabs}
\usepackage{threeparttable}
\usepackage{tabularx}
\usepackage{multirow}
\usepackage{orcidlink}
\usepackage{tabularray}
\usepackage{ragged2e}
\usepackage{circledsteps}
\pgfkeys{/csteps/inner color=white}
\pgfkeys{/csteps/fill color=black}
\usepackage{enumitem}
\usepackage{footmisc}
\makeatletter
\g@addto@macro\UrlSpecials{\do\!{\newline}}
\usepackage{acronym}
\acrodef{ML}[ML]{Machine Learning}
\acrodef{SOTA}[SOTA]{State-Of-The-Art}
\acrodef{MAC}[MAC]{Multiply-Accumulate}
\acrodef{QoS}[QoS]{Quality-of-Service}
\acrodef{CNN}[CNN]{Convolutional Neural Network}
\acrodef{RNN}[RNN]{Recurrent Neural Network}
\acrodef{TPU}[TPU]{Tensor Processing Unit}
\acrodef{AIMC}[AIMC]{Analog In-Memory Computing}
\acrodef{MVM}[MVM]{Matrix-Vector Multiplication}
\acrodef{PCM}[PCM]{Phase Change Memory}
\acrodef{FC}[FC]{Fully Connected}
\acrodef{CONV}[CONV]{Convolutional}
\acrodef{PU}[PU]{Processing Unit}
\acrodef{ADC}[ADC]{Analog-to-Digital Converter}
\acrodef{IMC}[IMC]{In-Memory Computing}
\acrodef{WL}[WL]{Word-Line}
\acrodef{BL}[BL]{Bit-Line}
\acrodef{DNN}[DNN]{Deep Neural Network}
\acrodef{RRAM}[RRAM]{Resistive Random-Access Memory}
\acrodef{SRAM}[SRAM]{Static Random-Access Memory}
\acrodef{DAC}[DAC]{Digital-to-Analog Converter}
\acrodef{ADC}[ADC]{Analog-to-Digital Converter}
\acrodef{HWA}[HWA]{Hardware-Aware}
\acrodef{SGD}[SGD]{Stochastic Gradient Descent}
\acrodef{NVM}[NVM]{Non-Volatile Memory}
\acrodef{FLMS}[FLMS]{First-Last Mapping Strategy}
\acrodef{LBMS}[LBMS]{Layer-Based Mapping Strategy}
\acrodef{FP}[FP]{Floating-Point}
\acrodef{DL}[DL]{Deep Learning}
\acrodef{SQuAD}[SQuAD]{Stanford Question Answering Dataset}
\acrodef{NLP}[NLP]{Natural Language Processing}

\title{\textsc{LionHeart}: A Layer-based Mapping Framework for Heterogeneous Systems with Analog In-Memory Computing Tiles}

\begin{document}
\author{
        Corey Lammie\orcidlink{0000-0001-5564-1356},~\IEEEmembership{Member,~IEEE,}\thanks{\hspace{-1em}\rule{3cm}{0.5pt} \newline \textcopyright  \hspace{1pt} 2025 IEEE. Personal use of this material is permitted. Permission from IEEE must be obtained for all other uses, in any current or future media, including reprinting/republishing this material for advertising or promotional purposes, creating new collective works, for resale or redistribution to servers or lists, or reuse of any copyrighted component of this work in other works.}
        Yuxuan Wang\orcidlink{0009-0008-4281-2388},~\IEEEmembership{Student Member,~IEEE,}
        Flavio Ponzina\orcidlink{0000-0002-9662-498X},~\IEEEmembership{Member,~IEEE,}
        \\
        Joshua Klein\orcidlink{0000-0002-4958-9990},~\IEEEmembership{Member,~IEEE,}
        Hadjer Benmeziane\orcidlink{0000-0002-5259-0749},~\IEEEmembership{Member,~IEEE,}
        Marina Zapater\orcidlink{0000-0002-6971-1965},~\IEEEmembership{Member,~IEEE,}\\
        Irem Boybat\orcidlink{0000-0002-4255-8622},~\IEEEmembership{Member,~IEEE,}
        Abu Sebastian\orcidlink{0000-0001-5603-5243},~\IEEEmembership{Fellow,~IEEE,}\\
        Giovanni Ansaloni\orcidlink{0000-0002-8940-3775},~\IEEEmembership{Member,~IEEE,}
        David Atienza\orcidlink{0000-0001-9536-4947}\thanks{
        
        \textit{Corresponding authors: I. Boybat and G. Ansaloni}
        
        Corey Lammie, Hadjer Benmeziane, Irem Boybat, and Abu Sebastian are with IBM Research - Zurich. email: ibo@zurich.ibm.com

        Flavio Ponzina, Yuxuan Wang, Giovanni Ansaloni, and David Atienza are with the Embedded Systems Lab, École Polytechnique Fédérale (EPFL). Joshua Klein, previously with EPFL, is currently with the Systems Integration Department of imec vzw in Leuven, Belgium. 
        Flavio Ponzina, previously with EPFL, is currently with the Computer Science and Engineering department at the University of California San Diego (UCSD). 
        email: giovanni.ansaloni@epfl.ch

        Marina Zapater is with the School of Engineering and Management Vaud (HEIG-VD) in the University of Applied Sciences Western Switzerland
        },~\IEEEmembership{Fellow,~IEEE}
        }
\markboth{IEEE Transactions on Emerging Topics in Computing}%
	{Lammie \MakeLowercase{\textit{et al.}}: \textsc{LionHeart}}
	\maketitle

\begin{abstract}
When arranged in a crossbar configuration, resistive memory devices can be used to execute \acp{MVM}, the most dominant operation of many \ac{ML} algorithms, in constant time complexity. Nonetheless, when performing computations in the analog domain, novel challenges are introduced in terms of arithmetic precision and stochasticity, due to non-ideal circuit and device behaviour. Moreover, these non-idealities have a temporal dimension, resulting in a degrading application accuracy over time. Facing these challenges, we propose a novel framework, named \emph{LionHeart}, to obtain hybrid analog-digital mappings to execute \ac{DL} inference workloads using heterogeneous accelerators. The accuracy-constrained mappings derived by \emph{LionHeart} showcase, across different \acfp{CNN} and one transformer-based network, high accuracy and potential for speedup. The results of the full system simulations highlight run-time reductions and energy efficiency gains that exceed 6$\times$, with a user-defined accuracy threshold for a fully digital floating point implementation.
\emph{LionHeart} is open-sourced here: \url{https://github.com/IBM/lionheart}.
\end{abstract}


\section{Introduction}
\IEEEPARstart{D}{eep} Learning (DL)-based solutions have been applied in various industrial and scientific applications, including computer vision, natural language processing, and speech recognition. 
Nowadays, \ac{SOTA} \ac{DL} models require significant memory to store their parameters (i.e., weights and biases), and billions of \ac{MAC} operations per inference. These extreme computing and memory requirements pose a challenge to deploy \ac{DL} workloads on edge devices, whose form factor and energy budget require limited computing capabilities and memory size.
However, the execution of \ac{DL} workloads at the edge represents a very attractive prospect, as it would allow the processing of data samples close to where it is collected, resulting in several benefits, including reduced latency, a higher degree of security and privacy, improved bandwidth efficiency, as well as increased resiliency. To enable \ac{DL} within the tiny resource envelopes of edge devices, workload optimization is key. To this end, many research works have proposed dedicated hardware solutions, customized for \ac{DL} workloads~\cite{Chen2020,Mehonic2020}.

\begin{figure}[!t]
    \centering
    \includegraphics[width=1\linewidth]{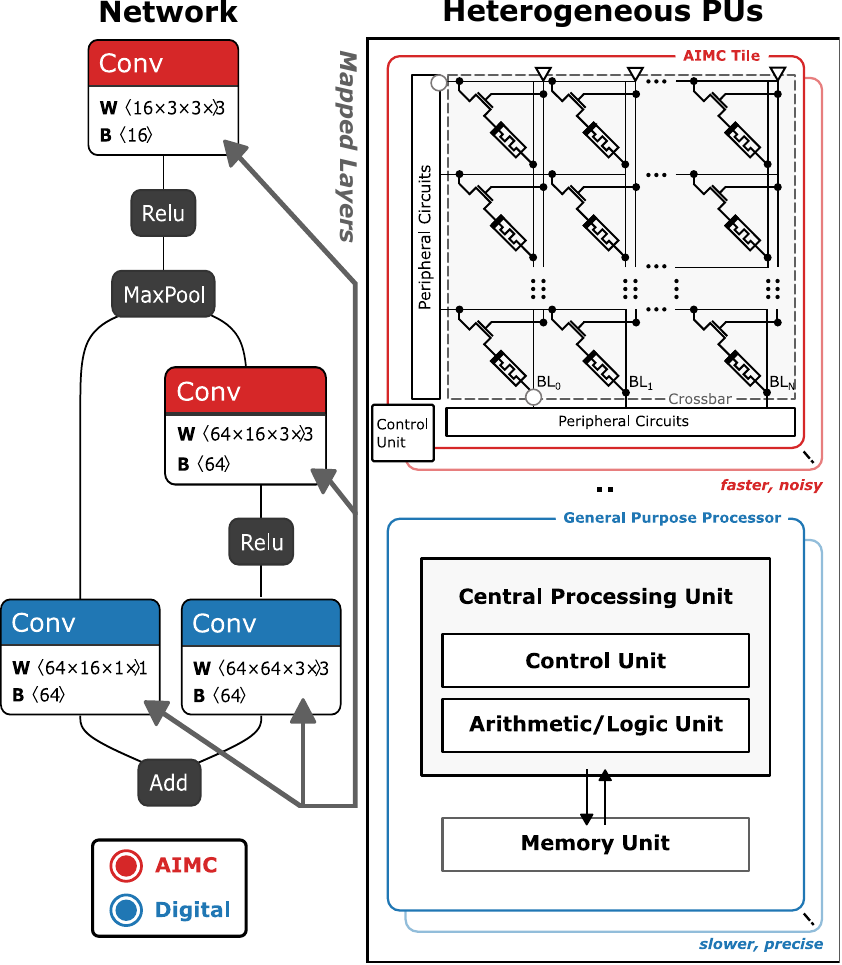}
    \caption{
    \emph{LionHeart} heterogeneously maps layers of ML networks to digital or analog resources, maximizes performance by exploiting \ac{AIMC} acceleration, while at the same time abiding to accuracy constraints.
    }
    \label{fig:key_difference_analog_digital}
\end{figure}

Such accelerators leverage the structured organization of \ac{ML} algorithms, such as \acp{DNN} and transformers, which are composed of layers sequences that, for the most part, implement linear algebra computing kernels. In turn, these can be effectively supported by hardware architectures by exploring their parallelism, whether employing reprogrammable logic or fixed-function implementations, such as \acp{TPU}~\cite{tpu_accelerators_review}. 
A disruptive approach in this space is \ac{IMC}, for which storage and computation is performed at the same physical location, avoiding the well-known memory wall problem. 

Different \ac{IMC} approaches have been proposed in the literature, operating at varying levels of the memory hierarchy and leveraging different technologies. 
We summarize this landscape in Section~\ref{sec:background}.
Among them, \acf{AIMC} is emerging as a particularly appealing alternative.
The \ac{AIMC} computing paradigm is based on crossbar architectures, with devices having variable conductance connecting rows and columns. By leveraging Ohm's law and Kirchhoff's current law, \ac{AIMC} accelerators can be used to perform matrix-vector operations between an input vector (encoded using \ac{WL} voltages) and a matrix (encoded using conductance) in \emph{linear} time~\cite{alpine}, i.e., the time required to write inputs and read-back outputs.

An important drawback hampering the widespread adoption of \ac{AIMC} is its sensitivity to device and circuit-related variations such as temporal conductance variations, device stochasticity, and circuit non-idealities, which can adversely affect accuracy~\cite{Mehonic2020}.
Many ongoing efforts investigate these effects and propose strategies to mitigate them, including \ac{HWA} retraining~\cite{Gallo2023} and model co-optimization~\cite{analog_nas}. 
However, none of these works adequately addresses the accuracy challenge from a system perspective, i.e. how \ac{AIMC} acceleration can be leveraged to speed up ML execution while faithfully modeling and controlling the ensuing noise-induced accuracy degradation.

We herein aim at filling this gap. To this end, we present a novel framework that enables the effective exploration of hybrid digital/analog mappings of \ac{DL} models (Fig.~\ref{fig:key_difference_analog_digital}), taking into account the current limitations of \ac{AIMC} which affect accuracy. Our framework effectively navigates the trade-off between run-time, performance, and accuracy degradation. 
Recognizing the exponential relation between the number of \acp{MVM} that can be executed in either the digital or analog domain, and the possible mapping solutions, we herein introduce an accuracy-driven heuristic that, by considering the largest \ac{MVM} operations first (in terms of \ac{MAC} operations), reduces the exploration problem to that of linear complexity.
The main contributions presented in this work are summarized as follows:
\begin{itemize}
    \item We present a novel accuracy-driven training framework, able to explore the space of hybrid digital/analog implementations of \acp{DNN}. The framework allocates different layers to digital or analog resources to minimize run-time while constraining the adverse effect of analog computations on accuracy. 
    Additionally, it is hardware-agnostic, meaning it can be applied to \ac{AIMC} crossbars adopting different device technologies;
    \item We derive a set of optimized hybrid mappings ensuring user-defined accuracy levels for a number of \acp{DNN} trained for CIFAR-10 and CIFAR-100 image classification, in addition to a transformer-based network for \ac{SQuAD} -- a \ac{NLP} task;
    \item We showcase, that for all of these networks, a sizable share of their workload can be executed in the analog domain, even for degradation thresholds as low as 0.5\% with respect to floating-point precision implementations;
    \item We investigate the speedup-accuracy tradeoff achieved by our mappings. To this end, we perform power and performance evaluations on different types of ML workloads using a cycle-accurate simulator based on gem5-X \cite{alpine};
    \item We compare our optimal mappings to those obtained from prior work (DIANA~\cite{diana} and Harmonica~\cite{behnam2022algorithmhardware}), and evaluate the variability of the achieved mappings, in addition to their accuracy and \ac{MAC} ratio;
    \item For one network, we hardware-aware train all possible mappings and establish a upper-bound for comparison;
    \item Finally, we recognize and address the real-world system accuracy degradation caused by temporal variation of programmed \acp{AIMC} crossbars at a user-desired evaluation time, $t_{eval}$.
\end{itemize}

\section{Background and Related Work}\label{sec:background}

\subsection{In-Memory Computing}\label{sec:background_aimc}
Traditional von Neumann computing systems involve separate processing and memory units, which consume significant energy and introduce additional latency when data is moved between them.
\ac{IMC} is a non-von Neumann computational approach, in which computation and storage are performed at the same physical location.
Several different \ac{IMC} architectures, which aim to blend computation and storage, have emerged in recent years~\cite{Verma2019}.
Architectures can utilize \ac{IMC} at different levels of the memory hierarchy, for example, when interfacing main memory~\cite{SIMDRAM}, as part of smart caches~\cite{BLADE}, or as functional units that augment processor pipelines and register files~\cite{alpine}.
While we consider this last scenario in our experimental evaluation when evaluating speedups, our methodology is agnostic with respect to the integration choice, because we abstract architectural aspects when performing application mapping to analog/digital resource.
From a technology perspective, both SRAM- and DRAM-based \ac{IMC} architectures have been proposed \cite{Mehonic2020}. 
Computation in these architectures can be achieved by taking advantage of charge sharing when simultaneously activating multiple memory rows.
A promising alternative, which we focus on in this paper, is that of employing \ac{NVM} devices, organized in a crossbar arrangement \cite{Mehonic2020}. These devices, used to encode network weights in a differential mapping scheme, are employed at the junction of word- and bit-lines, and their resistances are modulated according to the target weight value at compile time.

\begin{figure}[!t]
    \centering\includegraphics[width=0.85\linewidth]{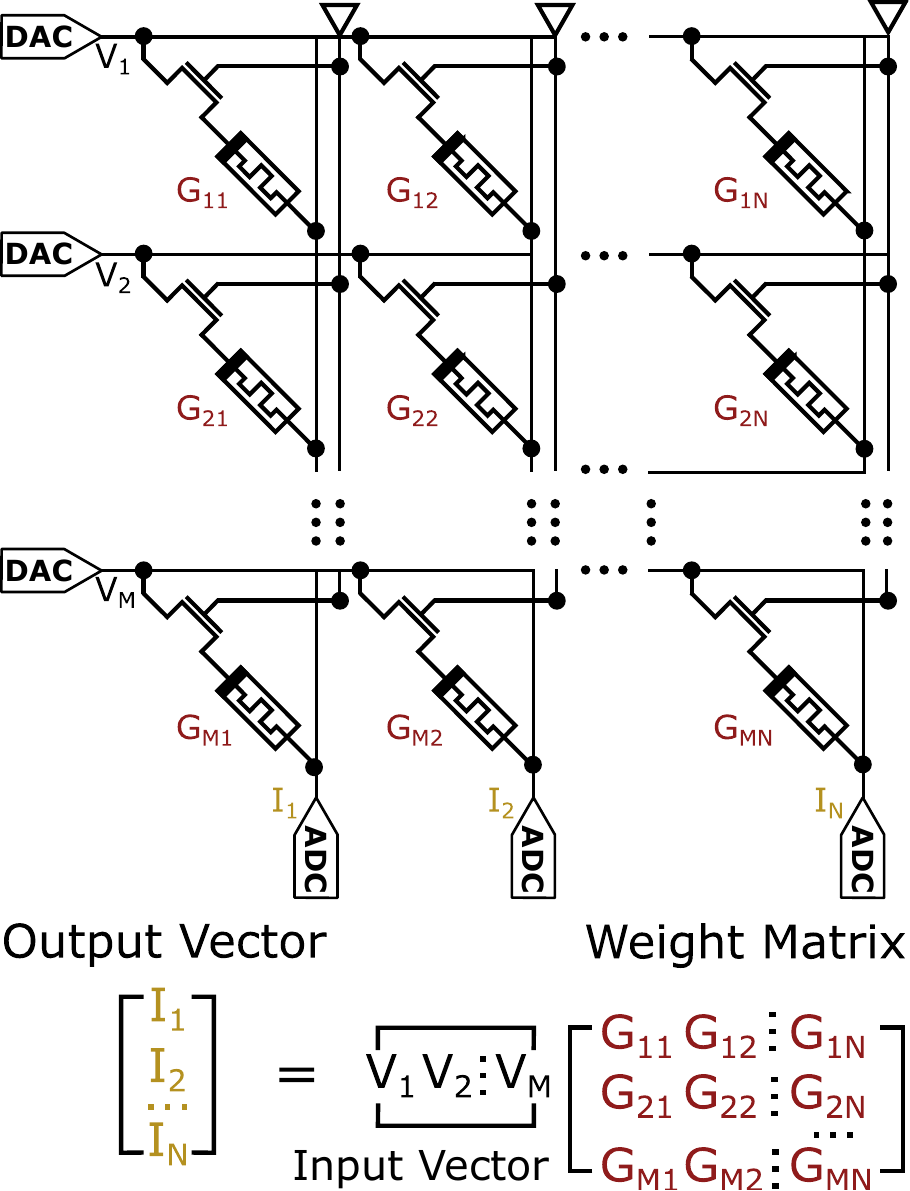}
    \caption{\textbf{(a)} Depiction of an \ac{AIMC} tile and its underlying compute mechanism. Each memristor, as depicted, is representative of a unit cell\protect\footnotemark.
    }
    \label{fig:AIMC_mechanism}
\end{figure}

As depicted in Fig.~\ref{fig:AIMC_mechanism}, \ac{AIMC} crossbars encode inputs as word line voltages using \acp{DAC}, and currents, representative of \acp{MVM} results, are collected across bit lines into \acp{ADC}.
Operations on \ac{AIMC} crossbars are performed in parallel, and hence, can be realized in constant time complexity.
For instance, a 256$\times$256 \ac{PCM} crossbar is shown to perform a total of 65,536 \ac{MAC} operations in 130ns~ in Le Gallo et al.\cite{LeGallo2023}.
However, as computation is performed in the analog domain, outputs are affected by non-deterministic noise, deviating from expected results. This is due to a number of device and circuit non-idealities, including temporal and temperature-dependent conductance variations, device stochasticity, \ac{ADC} mismatch, and IR drop~\cite{Gallo2023}.

\footnotetext{AIMC tiles may be realized using different architectures~\cite{Yu2021,Sun2023}, most prominently, pseudo-crossbar or conventional memory, where the input injection ways are different, and the single-bit-multi-bit encoding capabilities are also different. \textit{LionHeart} is architecture agnostic.}

\begin{figure}[!t]
    \centering
    \includegraphics[width=0.9\linewidth]{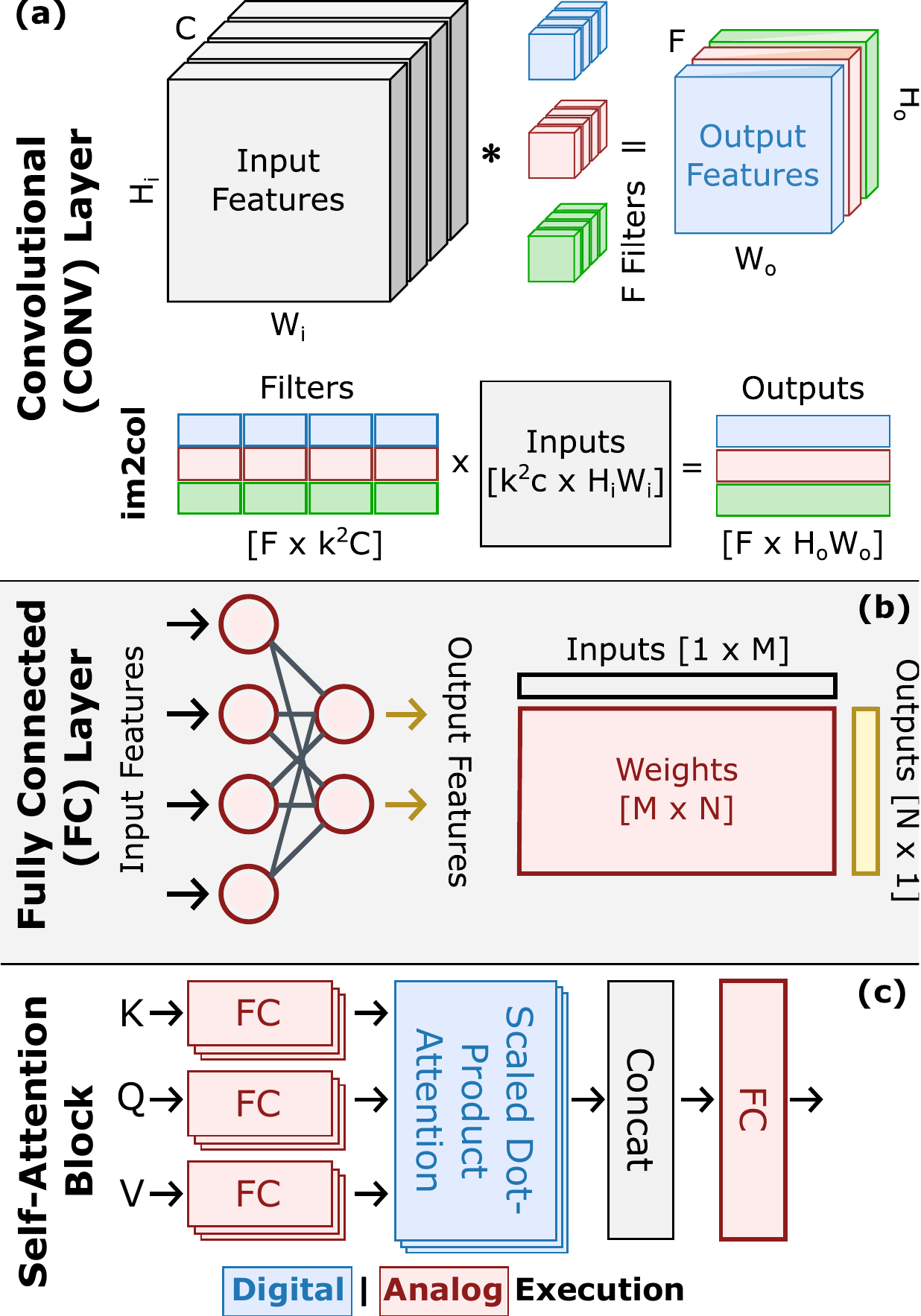}
    \caption{
    Mapping of \textbf{(a)} \ac{FC} and \textbf{(b)} \ac{CONV} layers to device conductances. Weights are linearly scaled and mapped between $G_{min}$ and $G_{max}$. 
    \textbf{(c)} Mapping/execution flow of a self-attention block when the maximum analog \ac{MAC} ratio is achieved.
    }
    \label{fig:conv_fc_layers}
\end{figure}

\begin{figure*}[!t]
    \centering
    \includegraphics[width = 1\linewidth]{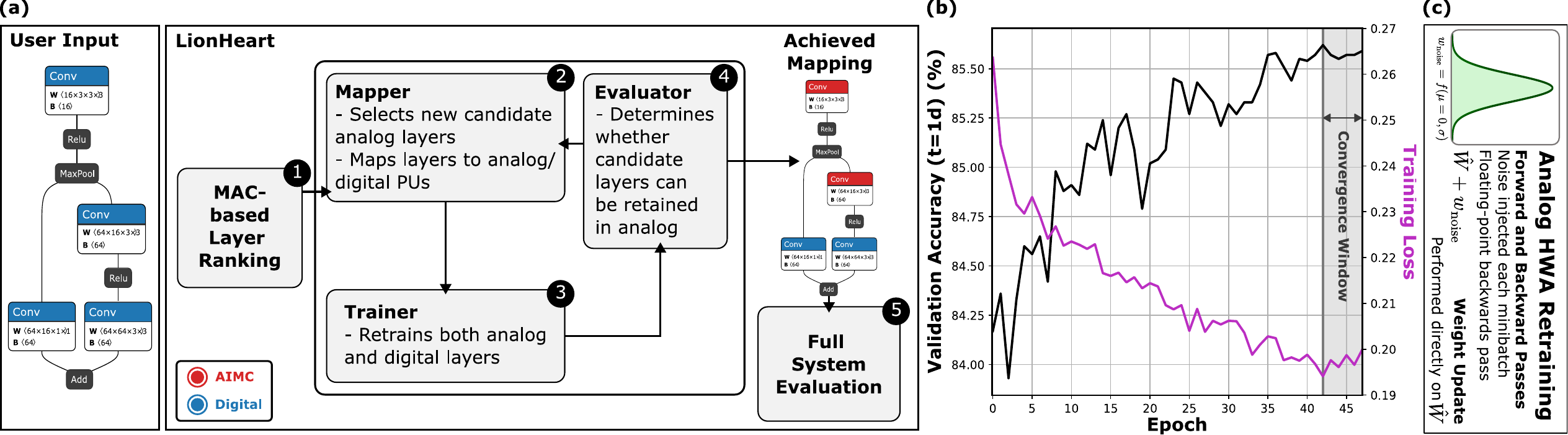}
    \caption{
\textbf{(a)} High-level overview of our proposed framework. 
\textbf{(b)} The validation accuracy at the desired evaluation time ($t_{eval}=$1d) and training loss of a candidate \ac{AIMC} layer during analog HWA retraining. Training is considered converged when, for a user-defined convergence window, the training loss does not decrease.
\textbf{(c)} High-level overview of analog HWA retraining.
}
    \label{fig:mapping_methodology}
\end{figure*}

\subsection{Hardware-Aware Training}
Analog \ac{HWA} offline training is a common approach to mitigate  the \ac{AIMC} device and circuit non-idealities, by making the model more robust for deployment.
When performing \ac{HWA} training, weight-noise is injected during forward propagation passes. Mathematically, this is expressed as follows: 
\begin{equation}
  \label{eq:mat-vec-analog}
    y_i = \alpha^\textrm{out}_i f_\textrm{adc}\big(\sum_j({w}_{ij} + \sigma_\textrm{w}\xi_{ij})(f_\textrm{dac}(x_j) + \sigma_\textrm{inp}\xi_j)  + \sigma_\textrm{out}\xi_i\big) + \beta_i,
\end{equation}
where $f_\text{adc}$ and $f_\text{dac}$ model the analog-to-digital and digital-to-analog processes, with dynamic scaling and range clipping, and $\xi$ denotes Gaussian noise. $\sigma$ indicates noise sampled from a zero-centered normal distribution. In addition to training noise-resilient weights, it is also possible to train other parameters, such as the input (\ac{DAC}) bounds of each tile, using back-propagation.  We employ the open-source AIHWKIT toolkit~\cite{Gallo2023} to expose these behaviours using system-level explorations in \emph{LionHeart}, allowing both the realistic characterization of analog non-idealities and analog HWA retraining.
While the toolkit, and by extension, the \emph{LionHeart} framework can be used to model other device technologies, such as \ac{RRAM}, in this paper, we model \ac{PCM}.
The \ac{PCM} device technology was chosen as compared to other device technologies, it utilizes fab–friendly materials and can achieve high conductance precision.
Section~\ref{sec:training_configuration} details the specific assumptions made during analog HWA retraining.

\subsection{Computational Complexity and Execution of Convolutional and FC Layers Using IMC}
\ac{FC} layers execute a number of \ac{MAC} operations which is linearly proportional to the number of input and output elements. Specifically, they have a complexity of $\mathcal{O}(MN)$, with $M$ and $N$ representing the number of input and output features, respectively. 
A \ac{CONV} layer that applies $F$ filters to compute an output feature map of size $W_o\times{}H_o\times{}F$ requires a number of MAC operations in the order of $\mathcal{O}(W_oH_oCK^2F)$, where $K$ represents the kernel size and $C$ the number of input channels. 

Weights of \ac{FC} layers can be computed directly using the \ac{MVM} operation, whereas weights of \ac{CONV} layers can be transformed to \acp{MVM} using the \emph{im2col} algorithm~\cite{im2col} (see Fig.~\ref{fig:conv_fc_layers}).
Using \emph{im2col}, weights of \ac{CONV} layers are unfolded and indexed as a matrix, where each column of the matrix contains the unrolled values of a  filter.
For both layer types, unfolded weight matrices are typically much larger than the attainable dimensions of \ac{AIMC} tiles. 
For both layer types, unfolded weight matrices are typically much larger than the attainable dimensions of AIMC tiles, e.g., all convolutional layers of ResNet20 require 32 256$\times$256 tiles.
Hence, \emph{tiling} can be employed to partition large matrices into a number of discrete tiles. This strategy requires an additional computing step to aggregate partial results, which is performed per tile, rather than per element~\cite{alpine}.
Unless otherwise stated, we utilize a distributed weight mapping scheme, in which larger matrix are \emph{evenly} distributed among tiles.

For self-attention blocks, we consider mapping the first three \ac{FC} layers of each head and the last \ac{FC} layer after concatenation.
\ac{FC} layers within the scaled-dot product attention are not considered, as they have \emph{dynamic} weights, which for \ac{AIMC} require constant reprogramming during infererence.

\subsection{Related Work}\label{sec:related_works}
Most related work, which performs application mapping on \ac{AIMC}-accelerated systems, consider resource limitations as their main (or only) constraint.
PUMA~\cite{PUMA} employs compiler analysis passes to define which parts of the application have the most potential for speedup, while AERO~\cite{AERO} formulates this same problem as a cost function minimization. 
Potential issues introduced by non-idealities and noisy environments, characterizing \ac{AIMC} devices, are overlooked in the above-mentioned works. They are similarly disregarded in~\cite{Song2017} and~\cite{Jin2021}, where all \ac{FC} and \ac{CONV} layers are executed using \ac{AIMC} tiles, without attempting to control induced accuracy degradation.
Some consideration of the effect of analog computing on accuracy is included in~\cite{diana}, where the authors propose a hybrid solution where the first and last layers of \acp{DNN} are executed on digital resources, postulating that these two layers are more sensitive to noise.

Two recent works present input channel-wise weight selection methods to map to heterogeneous analog-digital resources. 
The first, Harmonica~\cite{Behnam2022}, does so without employing any re-training strategy. The input is assumed to already be HWA trained, and some layers are confined to be implemented using digital resources. For channels of unconfined layers, the Hessian sensitivity~\cite{Dash2022} is determined. These are then ordered from the most sensitive to the least sensitive. The (next) most sensitive channel is converted to digital and the accuracy of the network is determined. This process is repeated until the accuracy of the network drops below a pre-defined threshold.
Harmonica is described in full in Algorithm 1 of~\cite{Behnam2022}.

The second, ODiMO~\cite{ODIMO}, relies on the unique approach of training a supernetwork, where for each channel, both digital and \ac{AIMC} implementations are trained simultaneously.
After the network is trained, for each channel, either the digital or \ac{AIMC} implementation is pruned.
Analog computation is modelled using 2-bit quantization.
Both of these methods do not consider (i) the variability of analog tiles as well as (ii) the temporal evolution of analog weights (conductances).

The complexity of devising high-performance, but accuracy-constrained mappings, is compounded by the varying size and sensitivity towards noise of the different layers in a DNN structure. Moreover, the number of possible mappings makes exhaustive exploration impractical, even for simple cases, as discussed, in the context of hardware quantization, in \cite{rios2023bit}. For example, even when layer-wise mapping is considered, a relatively low-depth model like AlexNet, composed of just seven layers, would therefore present $2^{7}$ possible analog-digital combinations. All of these would have to be trained and tested to arrive at an exact solution to the mapping problem. 
In contrast to Harmonica and ODiMO, we employ a heuristic-based layer-wise mapping strategy, which scales linearly with respect to the number of layers, as it only considers one target candidate layer at a time.

\section{\textsc{LionHeart} framework}

A high-level depiction of our framework is presented in Fig.~\ref{fig:mapping_methodology}. The input \ac{DNN} is assumed to be pre-trained in floating-point precision. 
Our methodology consists of: \Circled{1} a \ac{MAC}-based layer ranking (ordering) phase, in which the layers are indexed according to their computational requirements, \Circled{2} - \Circled{4} an optimization loop, evaluating layers greedily as possible candidates for \ac{AIMC} (analog) acceleration, and \Circled{5} a full system evaluation phase, where the full-system framework in~\cite{alpine} is used to estimate run time. In the remainder of this section, we detail these different components.

\noindent \textbf{MAC-based Layer Ranking} \Circled{1}: Ordering is performed by calculating the number of \ac{MAC} operations required by each layer. Note that this step only has to be performed once, as the number of \acp{MAC} only depends on the \ac{DNN} layer structure.

\noindent \textbf{Mapper} \Circled{2}: During the optimization loop, layers are considered, one at a time, in descending order with respect to their size (i.e., the number of required 
\acp{MAC}), and mapped to either analog or digital \acfp{PU}.
The rationale behind this approach is that larger layers have more redundancy and, hence, more leeway to compensate for the perturbations induced by analog computing \cite{analog_nas,Gallo2023}. Furthermore, the acceleration of a high number of \acp{MAC} with \ac{AIMC} computing has the potential to achieve higher speedups.
Initially, a selected layer is tentatively considered as \ac{AIMC}-accelerated. Its implementation is then transformed from its digital representation to an analog model provided by AIHWKIT, which accounts for non-idealities (noise, temporal drift, etc.).

\begin{table}[tp]
\centering
\caption{The number of layers, weights, and the top-1 (baseline, floating-point) accuracy of each evaluated network.}\label{tab:benchmarks}
\begin{threeparttable}
\resizebox{0.47\textwidth}{!}{%
\begin{tabular}{lrrrr}
\toprule \toprule
\textbf{Model} & \multicolumn{2}{>{\Centering\hspace{0pt}}m{0.33\linewidth}}{\textbf{Weights}} & \multicolumn{2}{>{\Centering\hspace{0pt}}m{0.309\linewidth}}{\textbf{Top-1 Acc./F1 (\%)}} \\ \midrule
(\# Mappable Layers) & CIFAR-10 & CIFAR-100 & CIFAR-10 & CIFAR-100 \\ 
\midrule
ResNet8   (10) & 77,360 & 77,968 & 87.38 & 59.40 \\
ResNet20  (20) & 268,336 & 270,522 & 90.52 & 65.15 \\
MobileNet (15) & 3,195,328 & 3,206,282 & 90.15 & 63.05 \\
VGG-16    (14) & 14,715,584 & 14,761,664 & 92.79 & 69.51 \\
AlexNet    (7) & 15,378,112 & 15,562,432 & 91.29 & 65.70 \\ 
\midrule
(\# Mappable Layers) & \multicolumn{2}{>{\Centering\hspace{0pt}}m{0.33\linewidth}}{SQuAD} & \multicolumn{2}{>{\Centering\hspace{0pt}}m{0.309\linewidth}}{SQuAD } \\ \midrule
MobileBERT (169) & \multicolumn{2}{>{\Centering\hspace{0pt}}m{0.33\linewidth}}{24,844,544\tnote{1}~} & \multicolumn{2}{>{\Centering\hspace{0pt}}m{0.309\linewidth}}{90.2~} \\
\bottomrule \bottomrule
\end{tabular}}
\begin{tablenotes}
\item[1] We use the following definition: \url{https://huggingface.co/docs/!transformers/model_doc/mobilebert}.
\end{tablenotes}
\end{threeparttable}
\end{table}

\noindent \textbf{Trainer} \Circled{3}: The transfer of a layer from digital to analog in the modeling environment usually has a large detrimental effect on accuracy. Nonetheless, most of the accuracy degradation at the desired evaluation time, $t_{eval}$, can be often countered with re-training. To this end, for each iteration, the entire network (both digital and analog layers) undergoes retraining, until no decrease in the loss function is observed inside a convergence window (see Fig.~\ref{fig:mapping_methodology}b). The width of the convergence window is a tuneable hyper-parameter of our framework, but can be small in practice (we set it to 5 epochs for the experiments in Section \ref{sec:expresults}) .

\noindent \textbf{Evaluator} \Circled{4}: The \ac{DNN} model accuracy after re-training with the achieved mapping is then compared to the user-specified drop threshold, i.e., the maximum drop between the floating-point and the average validation accuracy at the desired evaluation time.
If the optimized model adheres to this constraint, the tentative \ac{AIMC} acceleration of the layer, is confirmed. On the contrary, the optimization step is rolled back, and the layer is assigned for digital execution. The optimization loop proceeds by considering a further layer in its next iteration, and ends when all layers have been analyzed.

\noindent  \textbf{Full System Evaluation} \Circled{5}:
    After the achieved mapping has been determined, performance and energy statistics are obtained using the ALPINE framework~\cite{alpine}. ALPINE is a system simulator extending gem5-X~\cite{gem5} by providing hardware-validated models of AIMC accelerators. We employ it to collect performance statistics representative of AIMC-accelerated systems.

While not investigated in this paper, it is possible to extend the methodology of \textit{LionHeart} to co-apply different techniques to further improve accuracy, e.g., tunable ADC resolutions in RAELLA~\cite{Andrulis2023}.

\section{Evaluation}\label{sec:expsetup}
\subsection{Experimental Setup}

For the considered DNNs, similarly to \cite{diana} and \cite{behnam2022algorithmhardware} we evaluate the performance of \emph{LionHeart} on the CIFAR-10 dataset~\cite{Krizhevsky2009LearningML}. Moreover, we also provide outcomes on the more challenging CIFAR-100 dataset.~\cite{Krizhevsky2009LearningML}. We consider five \ac{DNN} architectures targeting edge devices, as listed in Table~\ref{tab:benchmarks}, where the size of the last \ac{FC} layer is modified according to the number of output classes (10 and 100, respectively).

While the degree to which a DNN architecture is parameterized for a given dataset is difficult to determine quantitatively, the DNN architectures were selected such that they range from being under- to over-parameterized so that they are representative of a broader range of networks and chosen tasks.
To demonstrate the generality of \emph{LionHeart} beyond CNNs, we also apply it to the MobileBERT transformer benchmark, processing on the SQuAD dataset~\cite{DBLP:journals/corr/RajpurkarZLL16}\footnote{We refer to the SQuAD 1.1 dataset as SQuAD throughout the paper.}.

We collect results using CIFAR-10 and CIFAR-100 for the first five \ac{DNN} architectures targeting edge devices as listed in Table~\ref{tab:benchmarks}.
Additionally, we collect results using SQuAD for MobileBERT.
The number of weights of these networks range from $\approx$78K to $\approx$25M.
Data is split into training, validation, and testing sets. For CIFAR-10 and CIFAR-100, the validation dataset was obtained by splitting the original training dataset into two separate datasets, comprising 90\% of training images and the remaining 10\% of validation images.
The same randomly sampled validation set was used for all experiments.
For SQuAD, the pre-existing development set (comprising 10\% of all inputs) was used as the validation set.
The test sets of all datasets were solely used for final model evaluation.

\begin{figure*}[!t]
    \centering
    \includegraphics[width = 1\linewidth]{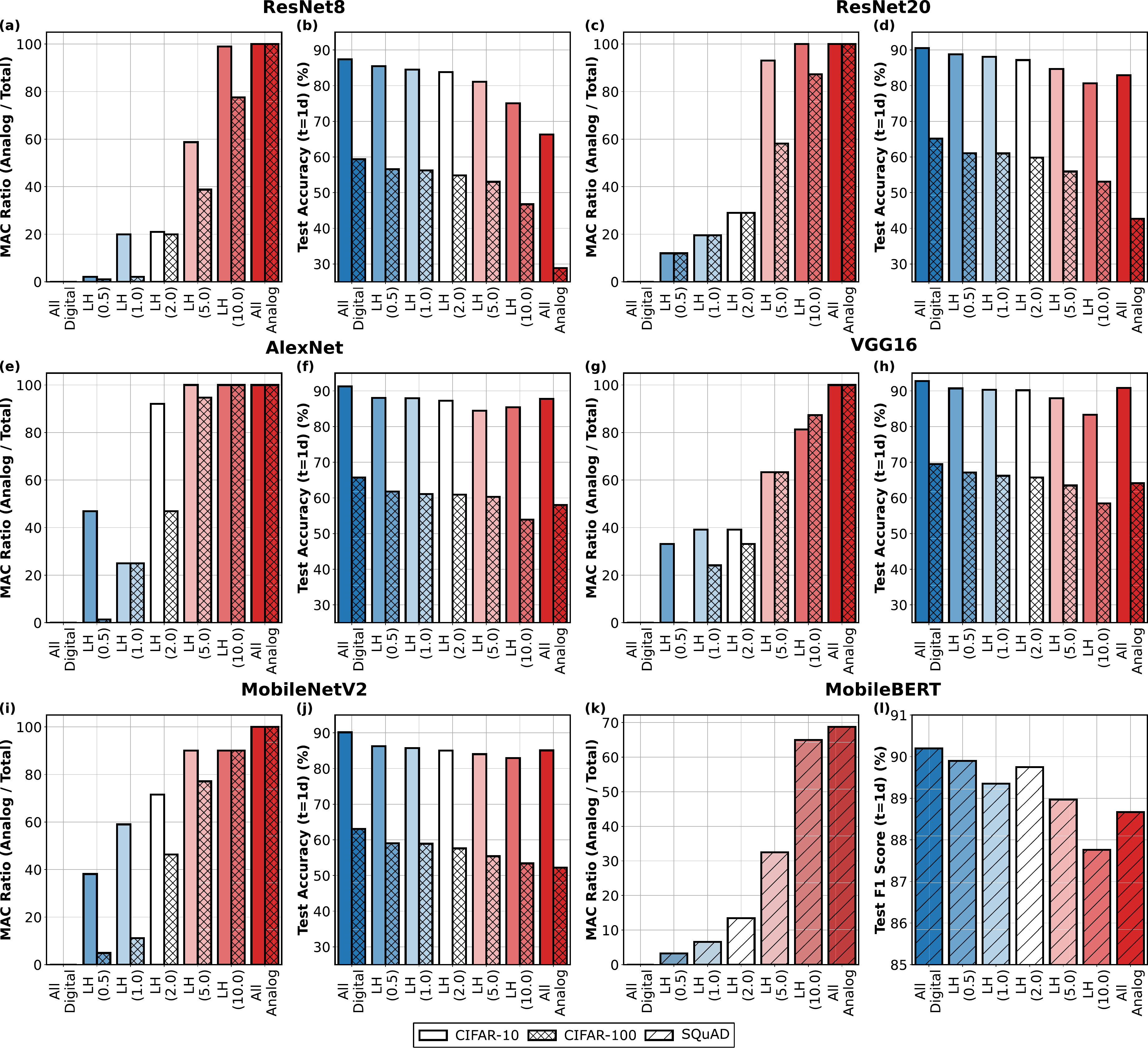}
    \caption{For (a,b) ResNet8, (c,d) ResNet20, (e,f) AlexNet, (g,h) VGG16, (i,j) MobileNetV2, and (k,l) MobileBERT, the \ac{MAC} ratio (as a percentage) and corresponding test set accuracies (F1 scores for SQuAD) at $t=$1d. The \ac{MAC} ratio quantity represents the ratio between the number of \acp{MAC} performed using \ac{AIMC} and the total number of \acp{MAC}}
    \label{fig:mac_ratio_plot}
\end{figure*}

\begin{table}[!t]
\centering
\caption{Shared training hyper-parameters for the CIFAR-10/ CIFAR-100 and SQuAD tasks.}\label{table:init_hyperparams}
\begin{tabular}{lcc}
\toprule \toprule
\textbf{Parameter} & \textbf{CIFAR-10/CIFAR-100} & \textbf{SQuAD} \\
\midrule
Optimizer & \multicolumn{2}{c}{SGD + Momentum} \\
Loss Function & \multicolumn{2}{c}{Cross Entropy} \\
LR Scheduler & CosineAnneling & Linear \\
T\_max & 50 & 3 \\
Batch Size & 256 & 8 \\ \midrule
\multicolumn{3}{c}{\textbf{Digital}} \\ \midrule
LR & 0.057 & 0.05 \\
Momentum & 0.867 & 0.99 \\
Weight Decay & 0 & 0.00054 \\ \midrule
\multicolumn{3}{c}{\textbf{Analog}} \\ \midrule
LR & 0.024 & 0.02 \\
Momentum & 0.775 & 0.9 \\
Weight Decay  & 0 & 0.003 \\
\bottomrule \bottomrule
\end{tabular}
\end{table}

\subsection{Training and Evaluation Strategies}
\subsubsection{Initial hyper-parameter exploration}
Our framework assumes that the input \ac{DNN} is pre-trained in floating-point precision. 
The selected models differ in depth, size, and complexity, and hence require different hyperparameters to achieve optimal training performance.
For all networks listed in Table~\ref{tab:benchmarks}, we empirically determined two sets of hyper-parameters, pertaining to the analog and digital domain, respectively. Then, these were applied to the networks. These are listed in Table~\ref{table:init_hyperparams}.
Further details on the adopted experimental setups in the two domains are provided in the following. 

\noindent \subsubsection{Digital training}
For all \acp{CNN}, networks were trained using floating-point precision until the training loss did not decrease for five consecutive epochs.
For MobileBERT, a fixed number (3) of training epochs was used.
All networks were trained with cross-entropy loss in conjunction with \ac{SGD} with momentum.

\noindent \subsubsection{Analog HWA re-training}~\label{sec:training_configuration}
To perform analog HWA re-training, the weights of the trained digital networks were employed as a starting point. The IBM AIHWKIT was used to inject noise during forward propagations. We refer the reader to~\cite{Gallo2023} for a comprehensive tutorial on \ac{HWA} training using IBM AIHWKIT.
To configure the behavior of \ac{AIMC} tiles and \ac{HWA} training, the IBM AIHWKIT utilizes different configurations which are defined using a standardized data class data structure.
A number of pre-defined configurations are provided.
We used the provided \texttt{InferenceRPUConfig()} configuration and \texttt{PCMLikeNoiseModel} phenomenological inference model.

The following additional modifications were made to the simulation configuration: (i) biases are assumed to be digital (i.e., they are not encoded on the last column of AIMC tiles). (ii) Channel- (i.e., column-) wise weight scaling, which has a negligible performance impact, is performed, where weights are mapped to a per-channel conductance range during learning after each mini-batch. (iii) The size of each \ac{AIMC} tile is assumed to be 256x256. (iv) Layers that span across multiple \ac{AIMC} tiles are assumed to be split evenly. (v) During training, multiplicative Gaussian noise is applied to unit weight values, with a standard deviation of $0.08$, extracted from hardware measurements provided by the authors of~\cite{LeGallo2023}.

\noindent \subsubsection{Accuracy evaluation}
As \ac{AIMC} hardware is inherently stochastic, multiple evaluation instances are required to determine representative performance metrics. Consequently, accuracy was measured 20 times and averaged across experiments to obtain the values reported in Section~\ref{sec:expresults}.  We considered five user-defined accuracy\footnote{For SQuAD, instead of considering the accuracy, the F1 score is considered as a percentage value.} thresholds of [0.5\%, 1.0\%, 3.0\%, 5.0\%, 10.0\%].
Moreover, analog \ac{NVM} devices such as \ac{PCM} are susceptible to temporal conductance drift. Since the behavior of \ac{AIMC} hardware can evolve over time, we explore in Section~\ref{sec:drift} accuracy metrics for \ac{AIMC} accelerators programmed considering drift levels predicted at [one second (i.e., $t=1s$), one minute (i.e., $t=60s$), one hour (i.e., $t=3600s$), one day (i.e., $t=86400s$)]. These are then evaluated at different points in time, to assess the effect of a mismatch between predicted and actual drift.

\noindent \subsubsection{Full System Evaluation}
The target simulated system for this work is a high-performance edge processor with an in-order ARM CPU core running at a frequency of 2.3GHz, with a 64KB each L1 data and instruction caches, 1MB of L2 cache, and 8GB of 2400MHz DDR4 RAM, as detailed in Table~\ref{tab:ALPINE-System-Specs}.
The software stack for the target \acp{DNN} includes the \acp{DNN} accelerated via the Eigen C++ library, running in user-space atop the Linux 5.4 kernel, and an Ubuntu 16.04 LTS disk image.  Each implementation resulting from a \emph{LionHeart} mapping is run for 10 inferences to reduce simulation noise. Measured execution time deviations among different runs of the same mappings never exceeded 4\% of the run-time. 

\subsection{Baselines}
\emph{LionHeart} (in all graphs, referred to as LH) is evaluated by comparing the accuracy and performance achieved with respect to two baselines: (i) \emph{Fully digital}. The floating-point \ac{DNN} model input of our methodology, where none of its layers is accelerated using \ac{AIMC}. It achieves the highest accuracy, but presents the longest run-time. (ii) \emph{Fully-analog}. A \ac{DNN} implementation executing all \ac{FC} and \ac{CONV} layers with static weights using \ac{AIMC}. This baseline represents the opposite of the \textit{fully-digital}: it maximizes performance, while potentially resulting in very large accuracy degradation.
In Section~\ref{sec:comparison_to_related_work}, we provide a further comparative evaluation with state of the art strategies, namely Ueyoshi et al. \cite{diana} (i.e., DIANA) and Harmonica \cite{behnam2022algorithmhardware}. 

\subsection{Accuracy Evaluation}\label{sec:expresults}

For five different drop thresholds and baseline configurations, we determined the ratio between \acp{MAC} performed using \ac{AIMC} and the total number of \acp{MAC} for CIFAR-10, CIFAR-100 and SQuAD (see Fig.~\ref{fig:mac_ratio_plot}). 
As expected, for all networks the \emph{All Digital} configuration has the highest test set accuracy and the \emph{All Analog} configuration has the lowest test set accuracy.
The proposed \emph{LionHeart} configurations exhibit a gradual transition from the accuracy of the largest to the smallest test set and a corresponding decrease in the digital/analog \ac{MAC} ratio between all configurations. 
Thus, it enables exploration of the performance/accuracy space exposed by \ac{AIMC} acceleration. Critically, the design space is navigated in a controlled way, that is, abiding to a user-defined accuracy degradation constraint. In this way, configurations exhibiting both high performance (high ratio of AIMC acceleration) and high accuracy can be derived. As an example, almost 60\% of the computation can be AIMC-accelerated in the ResNet-8 network for an accuracy threshold of 5\%, while an all-analog alternative would have a decrease in accuracy of 20\%. 

For a small number of networks and datasets, it is observed that the \emph{All Analog} configuration has a higher test set accuracy than some of the \emph{LionHeart} configurations; particularly those with larger drop thresholds, e.g., 10\%. Moreover, some \emph{LionHeart} configurations with larger drop thresholds have a higher accuracy than others with smaller drop thresholds.
To this end, we make the following observations.
First, heuristic-based methods are inherently stochastic in outcome. While we investigate the variability of \emph{LionHeart} and related work in Section~\ref{sec:comparison_to_related_work}, in this section, we first investigate the one-shot behavior of \emph{LionHeart} in earnest.
Second, \emph{LionHeart} enforces a \emph{global} drop threshold, meaning that if a network fails to converge when a layer is retrained using \ac{HWA} training, a large accuracy drop can be incurred (as large as this threshold).
This can result in a sub-optimal outcome if one of the first layers incurs a large accuracy drop, as all subsequent layers are subjected to the remaining drop threshold.

Lastly, given a fixed accuracy threshold, \emph{LionHeart} does not penalize configurations for being close to this threshold. This means that while in practice it would not make sense to specify, if an accuracy drop of X\% was indeed specified, and the
accuracy drop of the \emph{fully-analog} configuration was X-N\%, the resulting mapping would not be penalized for
being between X-N\% and X\%.

\begin{table}[!t]
\centering
\caption{Ratio of configurations where a layer is \ac{AIMC}-accelerated, per layer type and topological location in the MobileBERT encoder.}\label{table:average_retention_mobilebert}
\resizebox{\linewidth}{!}{%
\begin{tabular}{lccc}
\toprule \toprule
\textbf{Encoder Block Location} & \textbf{First} & \textbf{Middle (Avg.)} & \textbf{Last}      \\
\midrule
\textbf{Encoder Layer}  & \multicolumn{3}{c}{\textbf{Average Retention in Analog (\%)}}  \\
\midrule
Bottleneck                      & 45            & 52              & 46                 \\
Attention                       & 78             & 81              & 61                 \\
FFN0                            & 80             & 68              & 56                 \\
FFN1                            & 75             & 74              & 62                 \\
FFN2                            & 68             & 70              & 66                 \\
Intermediate                    & 48             & 75              & 48                 \\
Output                          & 45             & 52              & 40     \\
\bottomrule \bottomrule
\end{tabular}
}
\end{table}

We observe that no obvious pattern emerges regarding the topological order of layers in our explorations, other than that \ac{CONV} layers are more likely to be AIMC-accelerated compared to \ac{FC} ones. For the same layer type, whether or not a layer will be executed in the analog domain depends instead on the trained weights, and \emph{not} on the layers and the network topology, thus mandating exploration heuristics such as \emph{LionHeart}.
A similar conclusion can be reached for the MobileBERT case. Again, the choice of accelerated layers does not follow a clear topological pattern, as we report in Table~\ref{table:average_retention_mobilebert}, where the conversion percentage is reported for the first and last, in addition to the average of all middle encoder blocks.
The first attention layers are more amenable to be executed in AIMCs, but this is not the case for Bottleneck layers.

From our analysis, for both \acp{CNN} and transformer-based networks, we observe that the first layer is usually one of the most sensitive. The intermediate layers are generally seen to be more sensitive than the last layer. 
Our analyses demonstrate that, especially for larger networks with a more complex topology, properly mapping layers to digital or analog resources is non-trivial.

\begin{table}[!t]
    \centering\caption{
    Architectural parameters for full system simulation.
    }
    \fontsize{8pt}{8pt}\selectfont
    \begin{tabular}{lr}
        \toprule \toprule
        \textbf{CPU Model} & \textbf{1x In-Order @ 2.3GHz} \\ \midrule
        ISA & ARMv8 \\ 
        Supply Voltage VDD & 1.3V \\ 
        L1 Data/Instruction Cache & 64kB \\
        L2 Cache & 1MB \\
        Memory & 8GB DDR4 @ 2400MHz \\
        AIMC Tile Latency & 100ns \\
        AIMC Tile Data Transfer Bandwidth & 4GB/s \\
        AIMC Tile Energy Efficiency & 20 TOp/s/W \\ 
        CPU + SIMD (Baseline) Peak Efficiency & 22.8 GOp/s/W \\ \bottomrule \bottomrule
    \end{tabular}
    
    \label{tab:ALPINE-System-Specs}
\end{table}
\begin{figure}[!t]
    \centering
    \includegraphics[width=1\linewidth]{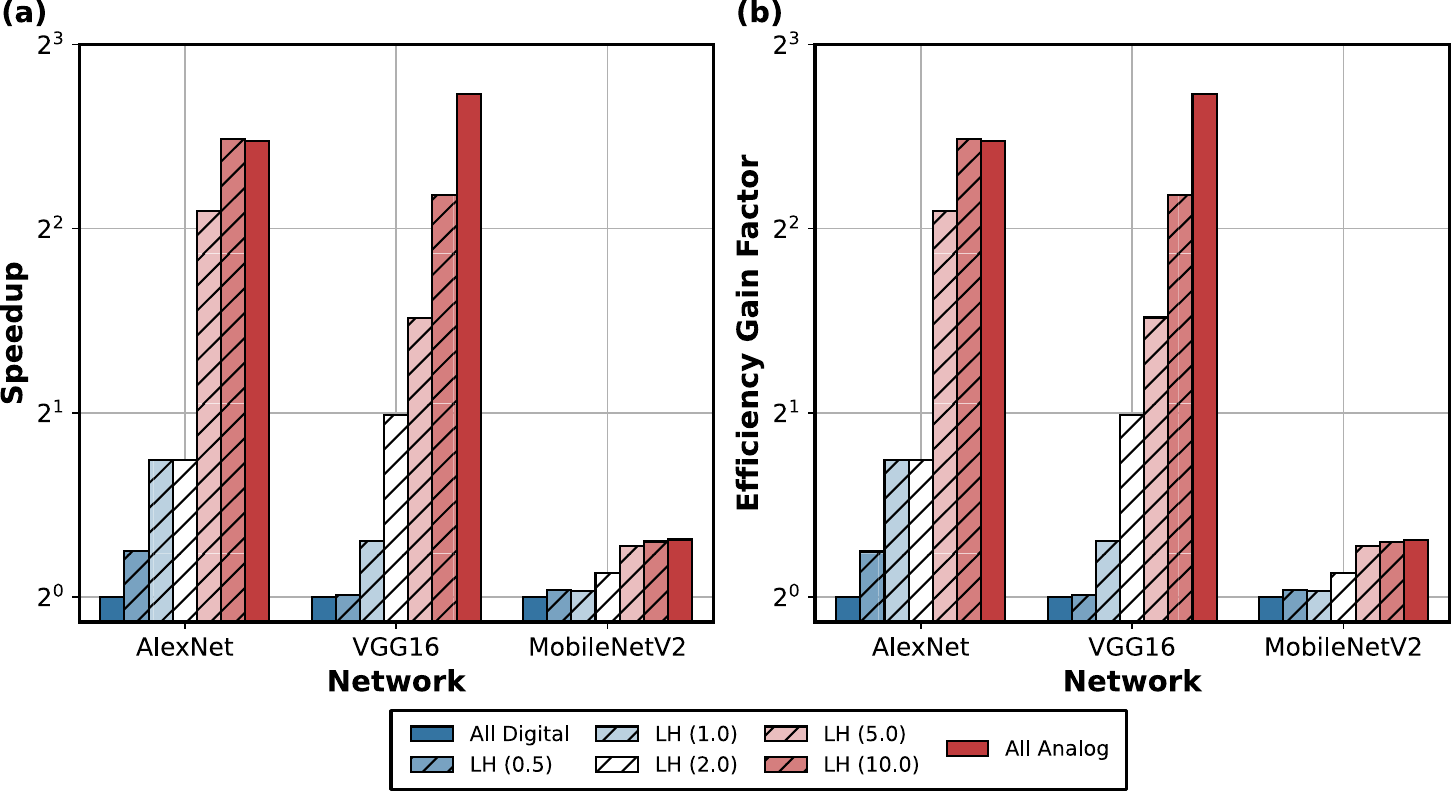}
    \caption{(a) Speedup and (b) energy efficiency gain resulting from different application mapping strategies, when performing an inference on benchmark \acp{DNN}. Data is normalized with respect to a fully digital execution with the ARM NEON SIMD co-processor in lieu of \ac{AIMC} acceleration.
    }
    \label{fig:energy_runtime}
\end{figure}

\subsection{Performance Evaluation}\label{sec:performance_eval}
In \emph{LionHeart}, the workload associated with \ac{DL} models and their layers are assumed to be architecture-agnostic, i.e., the ratio of analog \ac{MAC} to total \ac{MAC} operations is used as a proxy for speedup and energy efficiency.

This approach allows to derive hybrid mappings in a hardware-agnostic way, while also avoiding impractically lengthy simulations. On the other hand, it does not allow to directly quantify the performance and efficiency of mappings on a target architecture.
To inspect this aspect, and also validate the link between MAC ratio and performance/efficiency for one
system configuration, we herein  execute entire inferences for the three largest CNN benchmarks (AlexNet, VGG16 and MobileNetV2) and measure the ensuing run-times, including computations not pertaining to \ac{FC} nor \ac{CONV} layers. We investigate mappings derived by \emph{LionHeart}, with varying maximum accuracy degradation thresholds. 

As a test vehicle, we adopted a simulated system featuring a single-core in-order CPU clocked at 2.3GHz and a two-level cache hierarchy, modelled within the ALPINE framework~\cite{alpine}. 
The architecture embeds tightly-coupled \acp{AIMC}, governed by dedicated extensions to the ARM instruction set supported by the CPU. 
Table \ref{tab:ALPINE-System-Specs} lists the architectural parameters of the test system, which mirror those considered in~\cite{alpine} for the single-core, higher performance scenario.

To align the performance experiments with state-of-the-art low-power edge-domain devices, all \ac{DNN} implementations are quantized to INT8 precision.
The fully digital baseline architecture utilizes the ARM Neon SIMD co-processor in lieu of the tightly-coupled AIMC tiles. Note that due to the effects of memory transactions and latencies, the \textit{effective} Op/s of both the AIMC-enabled and baseline systems are a fraction of the peak Op/s, at 57\% and 35\%, respectively.

The CPU in the evaluated system architecture uses customized ISA extensions to access AIMC tiles that are tightly coupled to the CPU cores, without requiring the traversal of the memory hierarchy for data transmission~\cite{alpine}. The extension instructions are implemented using unused opcodes in the ARMv8 architecture. During inference, three instructions are used: \emph{CM\_QUEUE}, which places packed inputs in the input memory of an selected AIMC tile, \emph{CM\_PROCESS}, which operates the AIMC tiles and performs MVMs, and \emph{CM\_DEQUEUE}, which retrieves the outputs from selected AIMC tiles and places them in destination registers. For all speedup and efficiency gain factor evaluations, it is assumed that a single CPU core is used and the required number of tiles to map all analog weights for the evaluated network are tightly coupled to this core.

In Fig.~\ref{fig:energy_runtime}, we report the obtained speedups (with respect to fully-digital baselines) and efficiency gain factors. In order to align the performance experiments with state-of-the-art low-power edge-domain devices, all \ac{DNN} implementations are quantized to int8 precision. Additionally, the fully digital baseline architecture utilizes the ARM Neon SIMD co-processor in lieu of the tightly-coupled AIMC tiles. Note that due to the effects of memory transactions and latencies, the \textit{effective} Op/s of both the AIMC-enabled and baseline systems are a fraction of the peak Op/s, at 57\% and 35\% respectively.

These results highlight that the achievable speed-ups are highly dependent on the \ac{DNN} structure. In particular, even when all layers are computed in the analog domain, which is an upper bound for speedup and efficiency in our setting, only a speedup of 24\% can be reached in the MobileNetV2 benchmark. Such (comparatively) low run-time performance improvement is caused by the lower degree of data reuse of the depthwise separable convolutions employed in this network, which shift the balance between computation and memory access requirements toward the latter, limiting the benefit of accelerating \acp{MAC}. However, even in this case, \emph{LionHeart}, with a limited maximum accuracy degradation of 5\%, still captures almost all of the available speedup and energy efficiency gains.
\begin{figure*}[!t]
    \centering
    \includegraphics[width=1\linewidth]{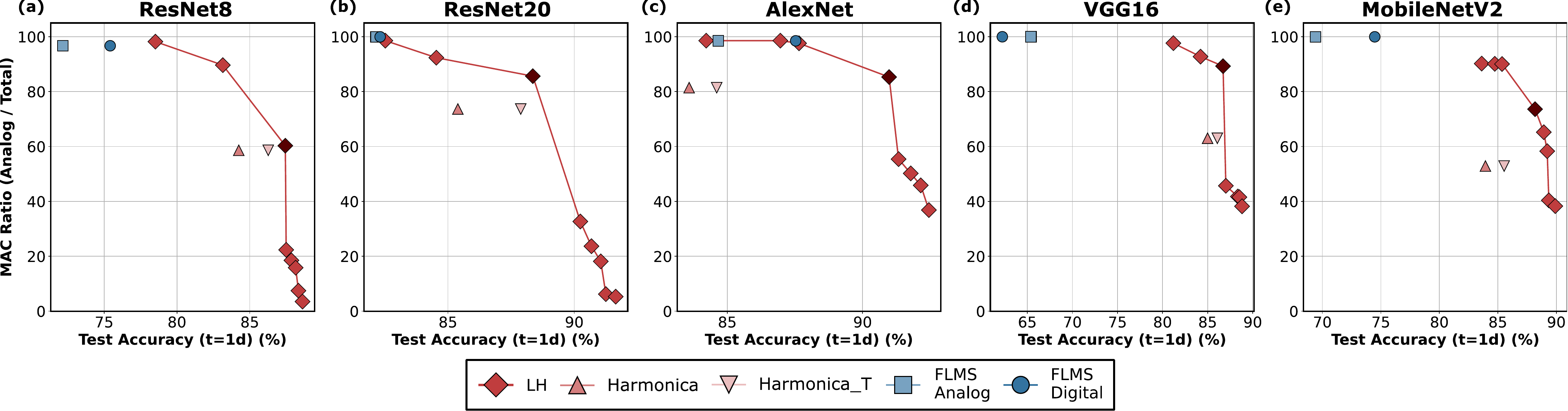}
    \caption{\emph{Approximate} Pareto front of the achieved \emph{LionHeart} mappings, and operational points of both variants of the Harmonica and FLMS methods. For each network, the elbow for \emph{LionHeart}, i.e., the optimal operating point, is highlighted using a darker hue value.
    }\label{fig:pareto_frontf}
\end{figure*}

AlexNet and VGG16, which do not employ separable convolutions, instead present more potential speedup (660\% in the full-analog case for VGG16, 550\% for AlexNet), due to the higher locality of the employed computations (which leads to less cache misses and memory-induced stalls). Even in those cases, \emph{LionHeart} can harness a large part of the potential speedup if enough accuracy leeway is allowed. For a 5\% accuracy degradation, a 425\% speedup was measured in AlexNet and a 285\% one in VGG16. AlexNet proved to be the most robust of the investigated networks. Finally, it can be noticed that system-level energy gains closely follow speedups. Indeed, \acp{AIMC} tiles are not a major contributor to overall system power (as opposed to processors and memories). Instead, they shorten run-time, ultimately lowering energy requirements. This is achieved both by allowing the execution of \acp{MVM} in constant time and by encoding DNN weights in the crossbar structure itself, hence easing the pressure on the memory hierarchy.

\begin{figure*}[!t]
    \centering
    \includegraphics[width = 1\linewidth]{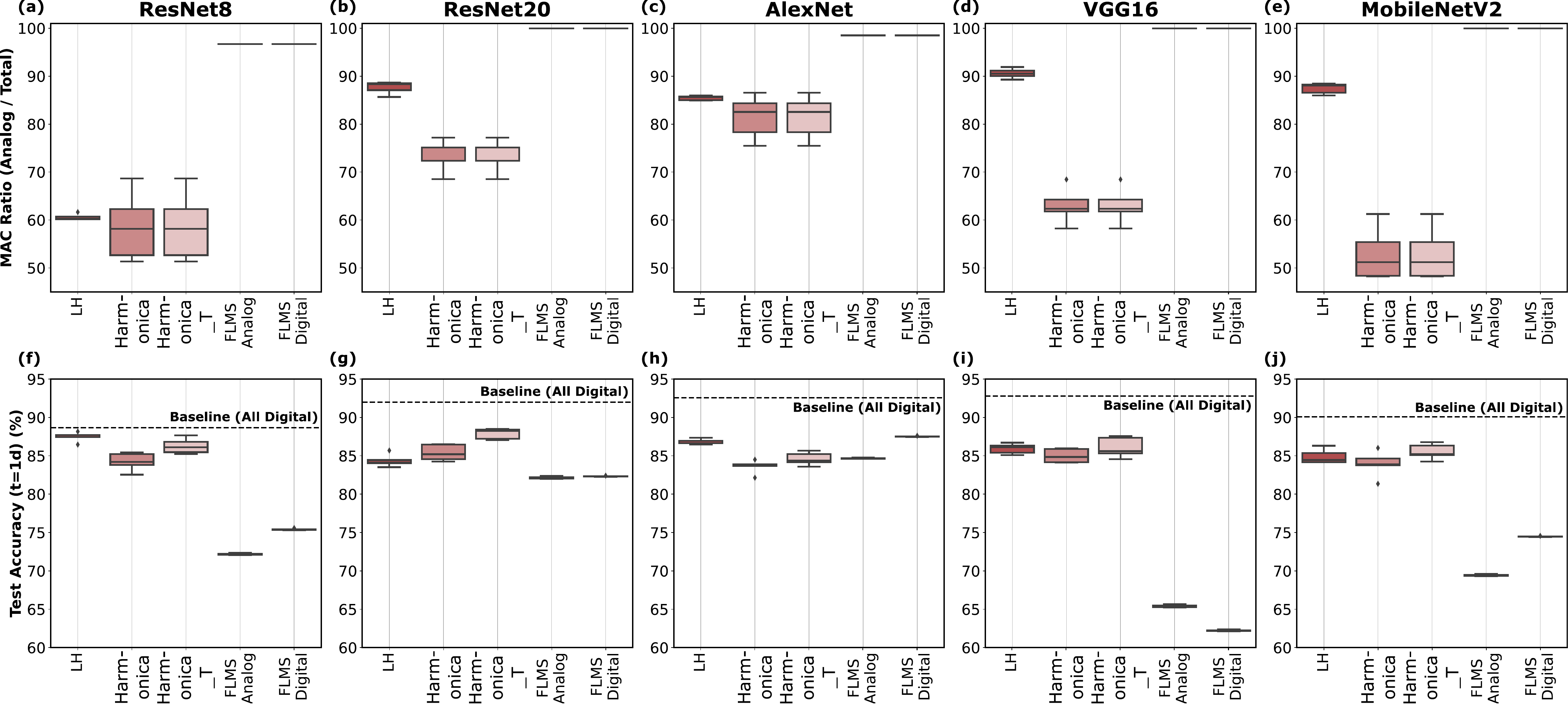}
    \caption{For the optimal LionHeart configuration, and both variations of the Harmonica and FLMS methods, (a-e) the \ac{MAC} ratio (as a percentage) between the number of \acp{MAC} performed using \ac{AIMC} and the total number of \acp{MAC}, and (f-j) the corresponding test set accuracies at $t=$1d.
    All experiments are repeated $n=5$ times.}
    \label{fig:related_work_mac_ratio_plot}
\end{figure*}

\begin{table}[!t]
\centering
\caption{Optimized hyper-parameters used for DIANA and Harmonica.}\label{table:hyperparameters}
\resizebox{\linewidth}{!}{%
\begin{tabular}{>{\hspace{0pt}}m{0.212\linewidth}>{\centering\hspace{0pt}}m{0.14\linewidth}>{\centering\hspace{0pt}}m{0.158\linewidth}>{\centering\hspace{0pt}}m{0.165\linewidth}>{\centering\hspace{0pt}}m{0.129\linewidth}>{\centering\arraybackslash\hspace{0pt}}m{0.131\linewidth}}
\toprule \toprule
\textbf{Parameter} & \textbf{ResNet8} & \textbf{ResNet20} & \textbf{MobileNet} & \textbf{VGG-16} & \textbf{AlexNet}  \\ \midrule
Optimizer          & \multicolumn{5}{>{\centering\arraybackslash\hspace{0pt}}m{0.9\linewidth}}{SGD + Momentum}     \\
Loss Function      & \multicolumn{5}{>{\centering\arraybackslash\hspace{0pt}}m{0.9\linewidth}}{Cross Entropy}      \\
LR Scheduler       & \multicolumn{5}{>{\centering\arraybackslash\hspace{0pt}}m{0.9\linewidth}}{CosineAnneling}     \\
T\_max             & \multicolumn{5}{>{\centering\arraybackslash\hspace{0pt}}m{0.9\linewidth}}{200}                \\
Batch Size         & \multicolumn{5}{>{\centering\arraybackslash\hspace{0pt}}m{0.9\linewidth}}{64}  \\ \midrule
\multicolumn{6}{>{\centering\arraybackslash\hspace{0pt}}m{0.935\linewidth}}{\textbf{Digital}}                        \\ \midrule
LR                 & 0.0945           & 0.00184           & 0.01866            & 0.0516          & 0.0157            \\
Momentum           & 0.86             & 0.99              & 0.94               & 0.88            & 0.9               \\
Weight Decay       & 0.00045          & 0.00017           & 0.00085            & 0.00058         & 0.00011           \\ \midrule
\multicolumn{6}{>{\centering\arraybackslash\hspace{0pt}}m{0.935\linewidth}}{\textbf{Analog}}                         \\ \midrule
LR                 & 0.0617           & 0.0882            & 0.0898             & 0.00619         & 0.00529           \\
Momentum           & 0.9              & 0.82              & 0.9                & 0.9             & 0.99              \\
Weight Decay       & 0.0008           & 0.00042           & 0.00037            & 0.00037         & 0.00091\\ \bottomrule \bottomrule  
\end{tabular}
}
\end{table}

\section{Comparison to Related Work}\label{sec:comparison_to_related_work}
We compare against two established heuristic-based methods, DIANA~\cite{diana} and Harmonica~\cite{behnam2022algorithmhardware}.
Comparisons are made using the CIFAR-10 dataset, as this dataset was originally used for evaluation by both of these methods.
For DIANA (referred to as \ac{FLMS} to encompass the more generic approach of executing only the first and last layer in high-precision), we consider two scenarios: (i) where the original network is in floating-point precision, akin to \emph{LionHeart}, and (ii) where the original network is \ac{HWA}-trained.

For Harmonica, we only consider layer-wise mappings, as channel-wise mappings are too computationally expensive, i.e., they require exploration of a much larger search space dependent on the number of channels rather than the number of layers, to be explored exhaustively.
Two scenarios are also considered: (i) where the mapped network is not retrained (Harmonica) and (ii) where the mapped network is retrained using \ac{HWA} retraining (Harmonica\_T).
The input to Harmonica is assumed to be \ac{HWA}-trained.
This is problematic, as \ac{HWA} training requires exhaustive hyper-parameter optimization to achieve optimal results~\cite{Gallo2023} when all layers are implemented using \ac{AIMC} hardware.
This is time demanding and computationally expensive.
We avoid comparisons with ODiMO due to their use of a channel-wise split, which markedly reduces the analog utilization, as evidenced by their lower analog mac ratio~\cite{ODIMO}.

To establish strong baselines for downstream comparison to \emph{LionHeart}, for each \ac{DNN} architecture, we performed hyper-parameter optimization for \ac{HWA} training using the Optuna~\cite{optuna} library. Specifically, we employed the Tree-structured Parzen Estimator (TPESampler) and the Median pruning algorithm (MedianPruner) optimization algorithms over 10 trials.
The resulting new hyper-parameters are listed in Table~\ref{table:hyperparameters}.

FLMS and Harmonica do not determine mappings for different \ac{MAC} ratio and accuracy values. Consequently, we determine the optimal \emph{LionHeart} mappings by generating an approximate Pareto front in Fig.~\ref{fig:pareto_frontf}.
Additionally, we determine and compare the \emph{average} operational point for each method (FLMS and Harmonica).
It is observed that the operational points of these methods are captured either \emph{on} or \emph{underneath} \emph{LionHeat's} Pareto front, meaning that \emph{LionHeart} discovers the same or even better mapping configurations along its front.

To perform further comparisons using a single operational point in Fig.~\ref{fig:related_work_mac_ratio_plot} for each method including \emph{LionHeart}, we select the elbow for each network.
The elbow method transverses operational points along the Pareto front from left-most point (where the quantity encoded using the y-axis is maximized) to the right-most point (where the quantity encoded using the x-axis is maximized). The optimal operation point, or elbow, is distinguished by the fact that before reaching it, the y-axis quantity remains almost unchanged, and after reaching it, rapidly decreases~\cite{Shi2021}.

\begin{figure}[!t]
    \centering
    \includegraphics[width = 0.95\linewidth]{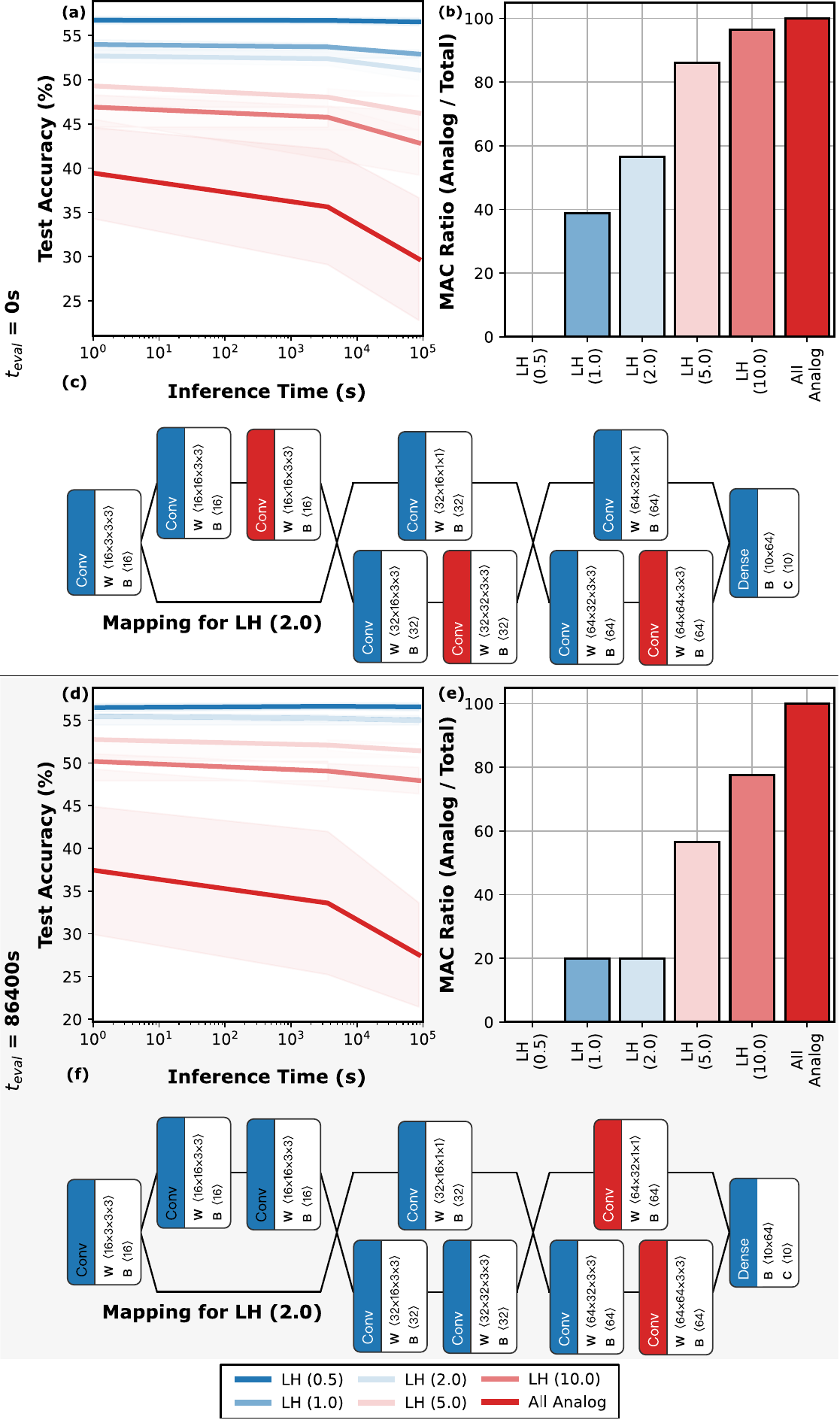}
    \caption{The impact of temporal drift when varying $t_{eval}$ in [0, 86400] for the ResNet8 network evaluated on CIFAR-100.}
    \label{fig:conv_fc_layers_exp}
\end{figure}

In Fig.~\ref{fig:related_work_mac_ratio_plot}, it can be observed that for all networks, both variants of \ac{FLMS} result in a high (near 100\%) \ac{MAC} ratio with zero variance. Although the resulting accuracy values have a small variance, they are \emph{unpredictable}, i.e., compared to other configurations, the average test accuracy varies greatly across networks.
Moreover, for most (4/5) networks, compared to other configurations, the accuracy values are the lowest.

On average, both variants of Harmonica yield higher accuracy values when compared to \ac{FLMS}, however, the average \ac{MAC} ratios are low and the variability of both the \ac{MAC} ratios and accuracy values are larger.
When the mapped networks from Harmonica are retrained using \ac{HWA} retraining, i.e., for Harmonica\_T, the average accuracy values increase. It is noted that these networks exhibit similarly variability to the networks originally generated from Harmonica.

\emph{LionHeart} consistently achieves competitive accuracy values with reduced variation compared to both Harmonica variants, for all networks expect ResNet20, while maintaining relatively high \ac{MAC} ratios.
For the single network (ResNet20) where LionHeart performs sub-optimally, it has a significantly higher \ac{MAC} ratio with respect to other configurations (not including FLMS). 
It is speculated that the reason Harmonica does not perform at least as well with respect to accuracy, is that at the beginning, in contrast to \textit{LionHeart}, it requires an already HWA retrained network, and does \emph{not} retrain any layers. \textit{LionHeart} begins with a stronger baseline (a pre-trained floating-point network) and performs retraining, providing an opportunity for accuracy losses to be recovered when each layer is converted to analog.

From Fig.~\ref{fig:pareto_frontf}, if another point along the Pareto front is taken with a smaller \ac{MAC} ratio, the resulting test precision significantly improves, that is, for a \ac{MAC} ratio of 85.63\%, the test accuracy is 88.36\%. This is larger than the average test accuracy for the other aforementioned configurations.

\section{Temporal Drift and Evaluation Time Analysis}
\label{sec:drift}
We study the impact of varying the evaluation time, $t_{eval}$, used during analog HWA retraining on the simulated inference time of the system.
In Fig.~\ref{fig:conv_fc_layers_exp}, these metrics are reported for all configurations (except \emph{All Digital}, which does not have a temporal dimension) using the ResNet8 network and CIFAR-100 dataset.
For both sets of mappings, mostly unique intermediate layers are retained in analog. Due to the (relatively) small drop threshold, the first and last layers for each are in digital.
It can be seen that the test accuracy decreases as a function of the drift time. Moreover, as the proportion of analog \acp{MAC} increases, the robustness to drift decreases.
Most notably, for the \emph{All Analog} configuration, independent of the evaluation time, the test accuracy decreases significantly ($>$10\%).
For all other configurations, no significant degradation is observed until $t>$1h. 
However, the temporal dependence of the test set accuracy is effected. Unsurprisingly, when a larger evaluation time is specified, the initial test set accuracy (at $t$=0) is also larger. While varied, for most configurations the robustness to drift is observed to improve as the evaluation time is increased.
This is significant because, when our methodology is used, a hard constraint on the average behavior of the drop in accuracy can be enforced at a given drift (evaluation) time.

\section{Comparison to Standard Hardware-Aware Retraining}

\begin{figure}[!t]
    \centering
    \includegraphics[width = 0.8\linewidth]{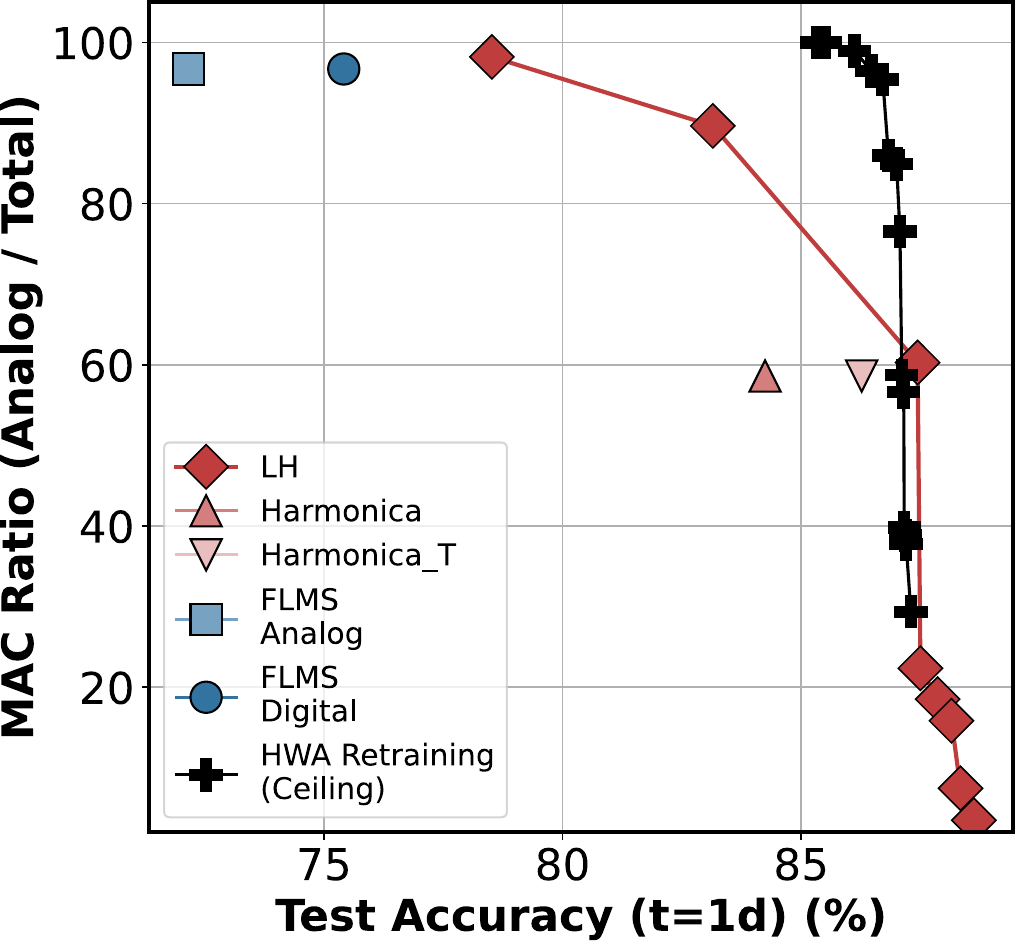}
    \caption{\emph{Approximate} Pareto front of the achieved LionHeart mappings, operational points of both variants of the Harmonica and FLMS methods, and the analog hardware-aware retraining ceiling for ResNet8 evaluated on CIFAR-10.}
    \label{fig:ceiling}
\end{figure}

Finally, to compare \emph{LionHeart} and other methods to standard hardware-aware retraining, we retrain all 1,024 mappings of ResNet8 and compute the exact Pareto front in Fig.~\ref{fig:ceiling}, which is indicative of the \emph{ceiling} or \emph{best case scenario} when all analog layers are retrained at the same time using the same convergence criteria for all other experiments.
It can be seen that for MAC ratio values smaller than 80\%, \emph{LionHeart} is extremely effective. For a MAC ratio of $\approx$60\%, \emph{LionHeart} exceeds this ceiling, indicating that progressively training analog layers can result in increased accuracy. This is in agreement with prior work~\cite{Rasch2023}.

\section{Conclusion}\label{sec:conclusions}
In this paper, we have presented \emph{LionHeart}, an accuracy-driven framework for mapping \ac{DL} workloads on hybrid analog-digital systems. The effectiveness of our method was demonstrated by deploying five different \acp{DNN} trained for image classification to heterogeneous \acp{PU}. Both the \ac{MAC} ratio between analog and digital operations and the effect of temporal drift were considered to enforce varying user-specified constraints. Performance evaluation was carried out for three \acp{DNN} using a simulated ARMv8-based processor with tightly coupled \ac{AIMC} tiles.
The optimal mappings of \emph{LionHeart} were compared to related work, and it was demonstrated that \emph{LionHeart} achieved state-of-the-art results.
Finally, additional analysis was performed by investigating the effects of temporal drift and the variation of the evaluation time.
Future work entails the hardware-based architecture-level exploration and investigation of the additional training time/complexity overhead that is incurred.

\section*{Acknowledgments}
This work was supported in part by the the Swiss NSF Edge-Companions project (GA No. 10002812); in part by the EC H2020 FVLLMONTI Project under Grant 101016776; in part by the ACCESS—AI Chip Center for Emerging Smart Systems, sponsored by InnoHK funding, Hong Kong, SAR; in part by the Swiss State Secretariat for Education, Research, and Innovation (SERI) through the SwissChips Research Project; and in part by EC H2020 WiPLASH under Grant 863337.

\bibliographystyle{IEEEtran}
\bibliography{EPFL_References}

\begin{thebibliography}{10}
\providecommand{\url}[1]{#1}
\csname url@samestyle\endcsname
\providecommand{\newblock}{\relax}
\providecommand{\bibinfo}[2]{#2}
\providecommand{\BIBentrySTDinterwordspacing}{\spaceskip=0pt\relax}
\providecommand{\BIBentryALTinterwordstretchfactor}{4}
\providecommand{\BIBentryALTinterwordspacing}{\spaceskip=\fontdimen2\font plus
\BIBentryALTinterwordstretchfactor\fontdimen3\font minus \fontdimen4\font\relax}
\providecommand{\BIBforeignlanguage}[2]{{%
\expandafter\ifx\csname l@#1\endcsname\relax
\typeout{** WARNING: IEEEtran.bst: No hyphenation pattern has been}%
\typeout{** loaded for the language `#1'. Using the pattern for}%
\typeout{** the default language instead.}%
\else
\language=\csname l@#1\endcsname
\fi
#2}}
\providecommand{\BIBdecl}{\relax}
\BIBdecl

\bibitem{Chen2020}
\BIBentryALTinterwordspacing
Y.~Chen, Y.~Xie, L.~Song, F.~Chen, and T.~Tang, ``A survey of accelerator architectures for deep neural networks,'' \emph{Engineering}, vol.~6, no.~3, pp. 264--274, 2020. [Online]. Available: \url{https://www.sciencedirect.com/science/article/pii/S2095809919306356}
\BIBentrySTDinterwordspacing

\bibitem{Mehonic2020}
X.~Yang, B.~Taylor, A.~Wu, Y.~Chen, and L.~O. Chua, ``Research progress on memristor: From synapses to computing systems,'' \emph{IEEE Transactions on Circuits and Systems I: Regular Papers}, vol.~69, no.~5, pp. 1845--1857, 2022.

\bibitem{tpu_accelerators_review}
A.~Shahid and M.~Mushtaq, ``{A Survey Comparing Specialized Hardware And Evolution In {TPUs} For Neural Networks},'' in \emph{2020 {IEEE} 23rd International Multitopic Conference ({INMIC})}.\hskip 1em plus 0.5em minus 0.4em\relax {IEEE}, Nov 2020.

\bibitem{alpine}
J.~Klein \emph{et~al.}, ``{ALPINE: Analog In-Memory Acceleration With Tight Processor Integration for Deep Learning},'' \emph{IEEE Trans. on Computers}, vol.~72, no.~7, 2023.

\bibitem{Gallo2023}
\BIBentryALTinterwordspacing
M.~Le~Gallo, C.~Lammie, J.~Büchel, F.~Carta, O.~Fagbohungbe, C.~Mackin, H.~Tsai, V.~Narayanan, A.~Sebastian, K.~El~Maghraoui, and M.~J. Rasch, ``{Using the IBM analog in-memory hardware acceleration kit for neural network training and inference},'' \emph{APL Machine Learning}, vol.~1, no.~4, p. 041102, 11 2023. [Online]. Available: \url{https://doi.org/10.1063/5.0168089}
\BIBentrySTDinterwordspacing

\bibitem{analog_nas}
H.~Benmeziane \emph{et~al.}, ``{AnalogNAS}: A neural network design framework for accurate inference with analog in-memory computing,'' in \emph{2023 IEEE International Conference on Edge Computing and Communications (EDGE)}, 2023.

\bibitem{diana}
K.~Ueyoshi \emph{et~al.}, ``{DIANA: An End-to-End Energy-Efficient Digital and ANAlog Hybrid Neural Network SoC},'' in \emph{2022 IEEE International Solid- State Circuits Conference (ISSCC)}, 2022.

\bibitem{behnam2022algorithmhardware}
P.~Behnam, U.~Kamal, A.~Shafiee, A.~Tumanov, and S.~Mukhopadhyay, ``Harmonica: Hybrid accelerator to overcome imperfections of mixed-signal dnn accelerators,'' in \emph{2024 IEEE International Parallel and Distributed Processing Symposium (IPDPS)}, 2024, pp. 619--630.

\bibitem{Verma2019}
N.~Verma, H.~Jia, H.~Valavi, Y.~Tang, M.~Ozatay, L.-Y. Chen, B.~Zhang, and P.~Deaville, ``In-memory computing: Advances and prospects,'' \emph{{IEEE} Solid-State Circuits Magazine}, vol.~11, no.~3, pp. 43--55, 2019.

\bibitem{SIMDRAM}
N.~Hajinazar, G.~F. Oliveira, S.~Gregorio, J.~D. Ferreira, N.~M. Ghiasi, M.~Patel, M.~Alser, S.~Ghose, J.~G{\'{o}}mez-Luna, and O.~Mutlu, ``{SIMDRAM}: a framework for bit-serial {SIMD} processing using {DRAM},'' in \emph{Proceedings of the 26th {ACM} International Conference on Architectural Support for Programming Languages and Operating Systems}.\hskip 1em plus 0.5em minus 0.4em\relax {ACM}, apr 2021.

\bibitem{BLADE}
\BIBentryALTinterwordspacing
W.~Simon, Y.~Qureshi, M.~Rios, A.~Levisse, M.~Zapater, and D.~A. Alonso, ``Blade: An in-cache computing architecture for edge devices,'' 2020. [Online]. Available: \url{https://www.semanticscholar.org/paper/e0af1fca1808230ac720257376697123c4bdf134}
\BIBentrySTDinterwordspacing

\bibitem{LeGallo2023}
\BIBentryALTinterwordspacing
M.~Le~Gallo \emph{et~al.}, ``{64-Core Mixed-Signal In-Memory Compute Chip Based on Phase-Change Memory for Deep Neural Network Inference},'' \emph{Nature Electronics}, vol.~6, no.~9, pp. 680--693, Jul 2023. [Online]. Available: \url{https://doi.org/10.1038/s41928-023-01010-1}
\BIBentrySTDinterwordspacing

\bibitem{Yu2021}
S.~Yu, H.~Jiang, S.~Huang, X.~Peng, and A.~Lu, ``Compute-in-memory chips for deep learning: Recent trends and prospects,'' \emph{IEEE Circuits and Systems Magazine}, vol.~21, no.~3, pp. 31--56, 2021.

\bibitem{Sun2023}
\BIBentryALTinterwordspacing
Z.~Sun, S.~Kvatinsky, X.~Si, A.~Mehonic, Y.~Cai, and R.~Huang, ``A full spectrum of computing-in-memory technologies,'' \emph{Nature Electronics}, vol.~6, no.~11, pp. 823--835, 2023. [Online]. Available: \url{https://doi.org/10.1038/s41928-023-01053-4}
\BIBentrySTDinterwordspacing

\bibitem{im2col}
A.~V. Trusov, E.~E. Limonova, D.~P. Nikolaev, and V.~V. Arlazarov, ``p-im2col: Simple yet efficient convolution algorithm with flexibly controlled memory overhead,'' \emph{{IEEE} Access}, vol.~9, pp. 168\,162--168\,184, 2021.

\bibitem{PUMA}
A.~Ankit \emph{et~al.}, ``{PUMA},'' in \emph{Proceedings of the Twenty-Fourth International Conference on Architectural Support for Programming Languages and Operating Systems}, 2019.

\bibitem{AERO}
S.~Yang \emph{et~al.}, ``{AERO: Design Space Exploration Framework for Resource-Constrained {CNN} Mapping on Tile-Based Accelerators},'' \emph{{IEEE} Journal on Emerging and Selected Topics in Circuits and Systems}, vol.~12, no.~2, pp. 508--521, 2022.

\bibitem{Song2017}
\BIBentryALTinterwordspacing
L.~Song, X.~Qian, H.~H. Li, and Y.~Chen, ``{PipeLayer: A Pipelined ReRAM-Based Accelerator for Deep Learning},'' in \emph{2017 IEEE International Symposium on High Performance Computer Architecture (HPCA)}, 2017. [Online]. Available: \url{https://www.semanticscholar.org/paper/6b3c06f148deba3926eff3c22ddf4dfd1195ac8a}
\BIBentrySTDinterwordspacing

\bibitem{Jin2021}
\BIBentryALTinterwordspacing
H.~Jin \emph{et~al.}, ``{ReHy: A ReRAM-Based Digital/Analog Hybrid PIM Architecture for Accelerating CNN Training},'' \emph{IEEE Trans. on Parallel and Distributed Systems}, 2021. [Online]. Available: \url{https://www.semanticscholar.org/paper/5ddbb41eaa61581af398a7a236bda28e5ecd0cee}
\BIBentrySTDinterwordspacing

\bibitem{Behnam2022}
\BIBentryALTinterwordspacing
P.~Behnam, U.~Kamal, and S.~Mukhopadhyay, ``An algorithm-hardware co-design framework to overcome imperfections of mixed-signal dnn accelerators,'' \emph{ArXiv}, vol. abs/2208.13896, 2022. [Online]. Available: \url{https://api.semanticscholar.org/CorpusID:251929459}
\BIBentrySTDinterwordspacing

\bibitem{Dash2022}
S.~Dash, Y.~Luo, A.~Lu, S.~Yu, and S.~Mukhopadhyay, ``Robust processing-in-memory with multibit reram using hessian-driven mixed-precision computation,'' \emph{IEEE Transactions on Computer-Aided Design of Integrated Circuits and Systems}, vol.~41, no.~4, pp. 1006--1019, 2022.

\bibitem{ODIMO}
M.~Risso, A.~Burrello, G.~M. Sarda, L.~Benini, E.~Macii, M.~Poncino, M.~Verhelst, and D.~J. Pagliari, ``Precision-aware latency and energy balancing on multi-accelerator platforms for dnn inference,'' in \emph{2023 IEEE/ACM International Symposium on Low Power Electronics and Design (ISLPED)}, 2023.

\bibitem{rios2023bit}
M.~Rios \emph{et~al.}, ``{Bit-Line Computing for CNN Accelerators Co-Design in Edge AI Inference},'' \emph{IEEE Trans. on Emerging Topics in Computing}, 2023.

\bibitem{gem5}
Y.~M. Qureshi \emph{et~al.}, ``{Gem5-X: A Gem5-Based System Level Simulation Framework to Optimize Many-Core Platforms},'' in \emph{2019 Spring Simulation Conference ({SpringSim})}.\hskip 1em plus 0.5em minus 0.4em\relax {IEEE}, 2019.

\bibitem{Andrulis2023}
\BIBentryALTinterwordspacing
T.~Andrulis, J.~S. Emer, and V.~Sze, ``Raella: Reforming the arithmetic for efficient, low-resolution, and low-loss analog pim: No retraining required!'' in \emph{Proceedings of the 50th Annual International Symposium on Computer Architecture}, ser. ISCA '23.\hskip 1em plus 0.5em minus 0.4em\relax New York, NY, USA: Association for Computing Machinery, 2023. [Online]. Available: \url{https://doi.org/10.1145/3579371.3589062}
\BIBentrySTDinterwordspacing

\bibitem{Krizhevsky2009LearningML}
\BIBentryALTinterwordspacing
A.~Krizhevsky, ``Learning multiple layers of features from tiny images,'' 2009. [Online]. Available: \url{https://api.semanticscholar.org/CorpusID:18268744}
\BIBentrySTDinterwordspacing

\bibitem{DBLP:journals/corr/RajpurkarZLL16}
\BIBentryALTinterwordspacing
P.~Rajpurkar, J.~Zhang, K.~Lopyrev, and P.~Liang, ``Squad: 100, 000+ questions for machine comprehension of text,'' \emph{CoRR}, vol. abs/1606.05250, 2016. [Online]. Available: \url{http://arxiv.org/abs/1606.05250}
\BIBentrySTDinterwordspacing

\bibitem{optuna}
\BIBentryALTinterwordspacing
T.~Akiba, S.~Sano, T.~Yanase, T.~Ohta, and M.~Koyama, ``Optuna: A next-generation hyperparameter optimization framework,'' in \emph{Proceedings of the 25th ACM SIGKDD International Conference on Knowledge Discovery \& Data Mining}, ser. KDD '19.\hskip 1em plus 0.5em minus 0.4em\relax New York, NY, USA: Association for Computing Machinery, 2019, p. 2623–2631. [Online]. Available: \url{https://doi.org/10.1145/3292500.3330701}
\BIBentrySTDinterwordspacing

\bibitem{Shi2021}
\BIBentryALTinterwordspacing
C.~Shi, B.~Wei, S.~Wei, W.~Wang, H.~Liu, and J.~Liu, ``A quantitative discriminant method of elbow point for the optimal number of clusters in clustering algorithm,'' \emph{EURASIP Journal on Wireless Communications and Networking}, vol. 2021, no.~1, p.~31, 2021. [Online]. Available: \url{https://doi.org/10.1186/s13638-021-01910-w}
\BIBentrySTDinterwordspacing

\bibitem{Rasch2023}
M.~J. Rasch \emph{et~al.}, ``{Hardware-Aware Training for Large-Scale and Diverse Deep Learning Inference Workloads Using In-Memory Computing-Based Accelerators},'' \emph{Nat. Communications}, vol.~14, no.~1, 2023.

\end{thebibliography}

\begin{IEEEbiography}[{\includegraphics[width=1in,height=1.25in,clip,keepaspectratio]{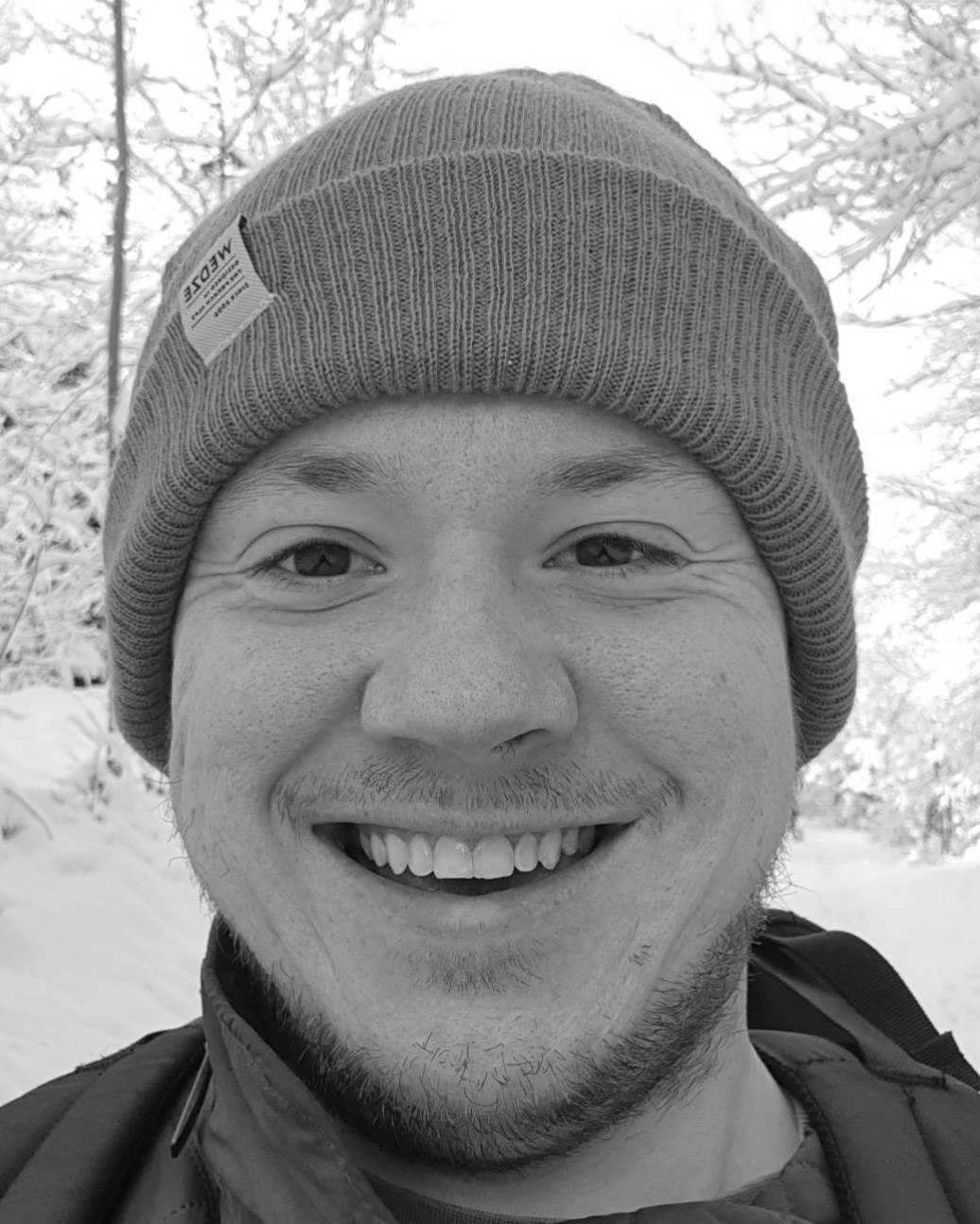}}]{Dr. Corey Lammie}
is a post-doctoral researcher in the IMC group at IBM Research - Zurich. He completed a PhD in Computer Engineering at James Cook University (JCU) in March, 2023, where he also completed his undergraduate degrees in Electrical Engineering (Honours) and IT, in 2018. He has received several awards and fellowships including the intensely competitive 2020-2021 IBM international PhD Fellowship, a Domestic Prestige Research Training Program Scholarship, and the 2020 CAS Society Pre-Doctoral Grant. Dr. Lammie has served as a guest editor for IEEE Journal on Emerging and Selected Topics in Circuits and Systems (JETCAS). In addition, he has served as a reviewer for several IEEE journals, and as a TC member for a number of conferences.
\end{IEEEbiography}

\newpage

\begin{IEEEbiography}[{\includegraphics[width=1in,height=1.25in,clip,keepaspectratio]{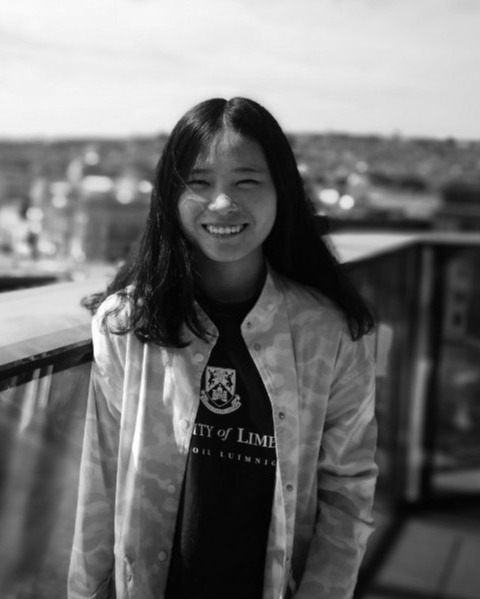}}]{Yuxuan Wang}
is a PhD student and doctoral assistant at the Embedded Systems Laboratory of the École Polytechnique Fédérale (EPFL), Switzerland, where she obtained her Master's degree in Electrical Engineering in 2023.
Her research interests include In-Memory Computing (IMC) and software-hardware co-optimization for efficient Deep Neural Network (DNN) acceleration.
\end{IEEEbiography}

\vspace{-4em}

\begin{IEEEbiography}[{\includegraphics[width=1in,height=1.25in,clip,keepaspectratio]{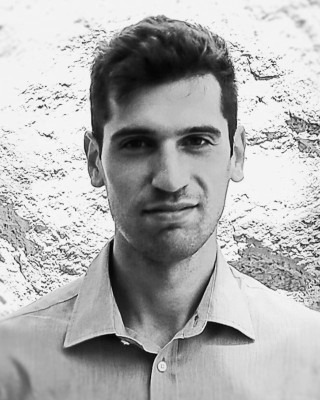}}]{Dr. Flavio Ponzina}
received the M.Sc. degree in Computer Engineering from Politecnico di Torino, Italy, in 2018, and the Ph.D degree in Electronic Engineering from EPFL, Switzerland, in 2023. He is currently a postdoctoral researcher at University of California, San Diego, United States. His main research interests include low power architectures and AI-based systems optimization.
\end{IEEEbiography}

\vspace{-4em}

\begin{IEEEbiography}[{\includegraphics[width=1in,height=1.25in,clip,keepaspectratio]{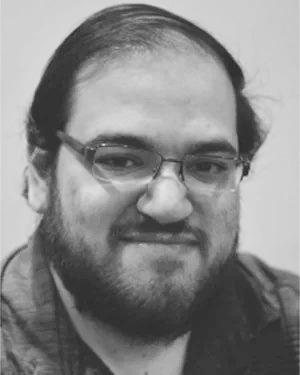}}]{Joshua Klein}
is a Researcher in the system integration department of imec, Belgium, specializing in modeling of near- and IMC accelerators and full systems. He previously worked in the ESL of the École Polytechnique Fédérale de Lausanne (EPFL), Switzerland, where he received his Ph.D. in Electrical Engineering in 2024. He received his B.Sc. in Computer Engineering in 2017 magna cum laude and his M.Sc. in Electrical and Computer Engineering in 2019 from Boston University, USA.  
\end{IEEEbiography}

\vspace{-4em}

\begin{IEEEbiography}[{\includegraphics[width=1in,height=1.25in,clip,keepaspectratio]{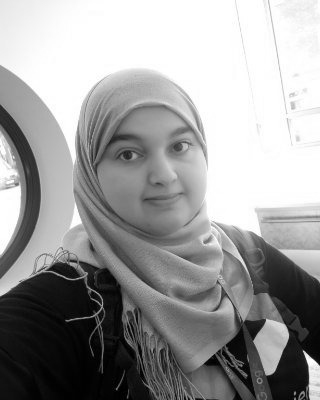}}]{Dr. Hadjer Benmeziane}
is an IBM researcher, specializing in hardware-aware NAS for emerging AI accelerators such as AIMC. She received her PhD from Université Polytechnique des Hauts-de-France in August 2023, following her Master's and Engineering degree in Computer Science from Ecole Supérieure d'Informatique, Algiers, Algeria. Her work on Analog NAS received the prestigious IEEE open source science award and best paper award at IEEE Services Computing 2023 Symposium. 
\end{IEEEbiography}

\vspace{-4em}

\begin{IEEEbiography}[{\includegraphics[width=1in,height=1.25in,clip,keepaspectratio]{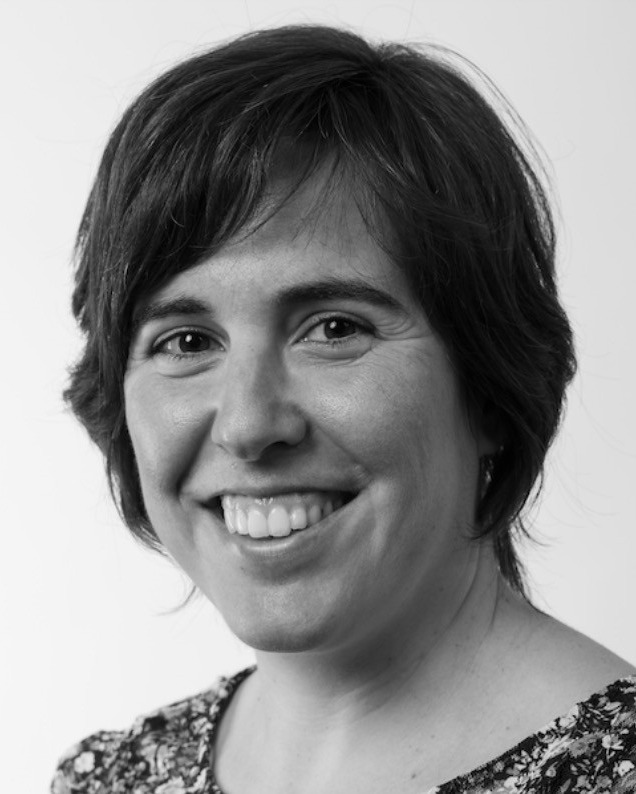}}]{Dr. Marina Zapater}
is an Associate Professor in the REDS Institute at the School of Engineering and Management of Vaud (HEIG-VD) of the University of Applied Sciences Western Switzerland (HES- SO) since 2020. She received her Ph.D. degree in Electronic Engineering from Universidad Polit´ecnica de Madrid, Spain, in 2015. Her research interests include thermal, power, and performance design and optimization of complex heterogeneous architectures.
\end{IEEEbiography}

\vspace{-4em}

\begin{IEEEbiography}[{\includegraphics[width=1in,height=1.25in,clip,keepaspectratio]{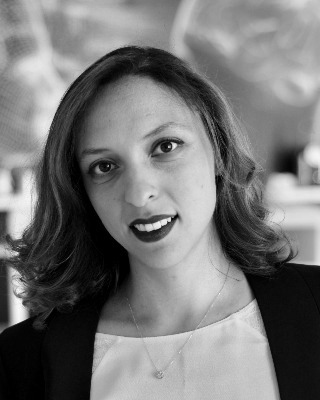}}]{Dr. Irem Boybat}
received the BSc degree in electronics engineering from Sabanci University, Turkey, in 2013, and the MSc and PhD degrees in electrical engineering from École Polytechnique Fédérale (EPFL), Switzerland, in 2015 and 2020, respectively. She is a research staff member with IBM Research – Zurich. Her research interests include in-memory computing for AI systems, neuromorphic computing, and emerging resistive memory.
\end{IEEEbiography}

\vspace{-4em}

\begin{IEEEbiography}[{\includegraphics[width=1in,height=1.25in,clip,keepaspectratio]{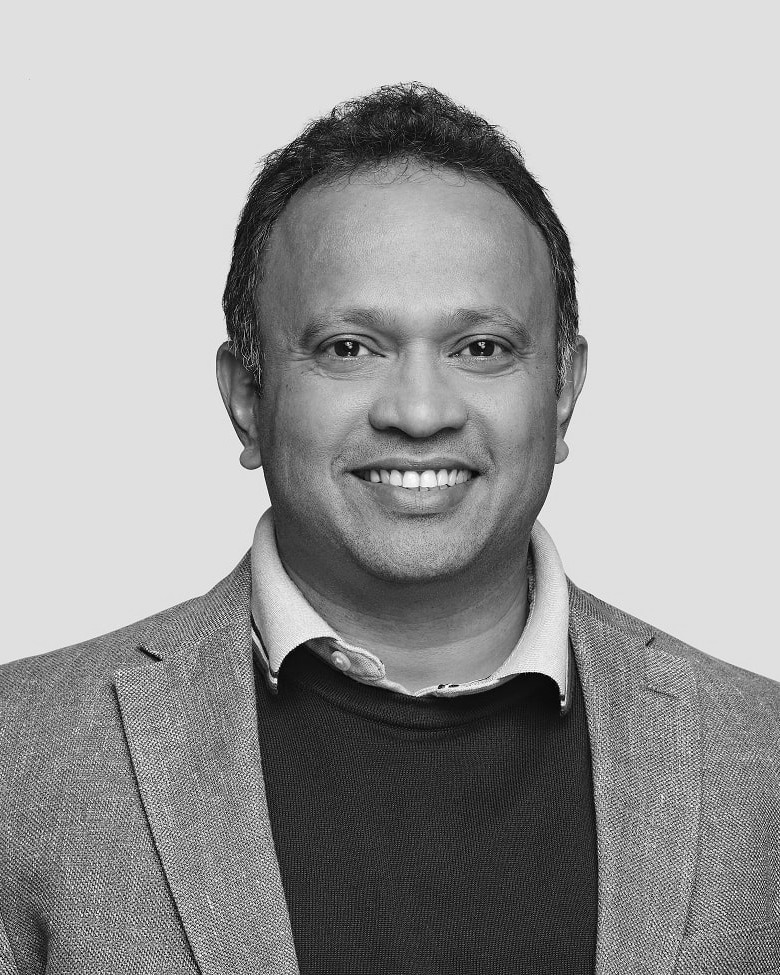}}]{Dr. Abu Sebastian}
is a Distinguished Research Scientist at IBM Research – Zurich. He received a B. E. (Hons.) degree in Electrical and Electronics Engineering from BITS Pilani, India, in 1998 and M.S. and Ph.D. degrees in Electrical Engineering (minor in Mathematics) from Iowa State University in 1999 and 2004, respectively. He was a contributor to several key projects in the space of storage and memory technologies and currently manages the research effort on IMC at IBM Research – Zurich. 
In 2015, he was awarded the European Research Council (ERC) consolidator grant and in 2020, he was awarded an ERC Proof-of-concept grant. He was elected an IBM Master Inventor in 2016. In 2019, he received the Ovshinsky Lectureship Award for his contributions to "Phase-change materials for cognitive computing" and in 2023, he was conferred the title of Visiting Professor in Materials by University of Oxford.
\end{IEEEbiography}

\vspace{-4em}

\begin{IEEEbiography}[{\includegraphics[width=1in,height=1.25in,clip,keepaspectratio]{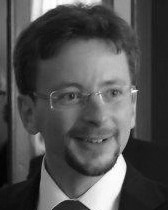}}]{Dr. Giovanni Ansaloni}
received the PhD degree in informatics from USI, in 2011. He is a researcher with the Embedded Systems Laboratory of EPFL (Lausanne, CH). He previously worked as a post-doc with the University of Lugano (USI, CH) between 2015 and 2020, and with EPFL between 2011 and 2015. His research efforts focus on domain-specific and ultra-low-power architectures and algorithms for edge computing systems, including hardware and software optimization techniques.
\end{IEEEbiography}

\vspace{-4em}

\begin{IEEEbiography}[{\includegraphics[width=1in,height=1.25in,clip,keepaspectratio]{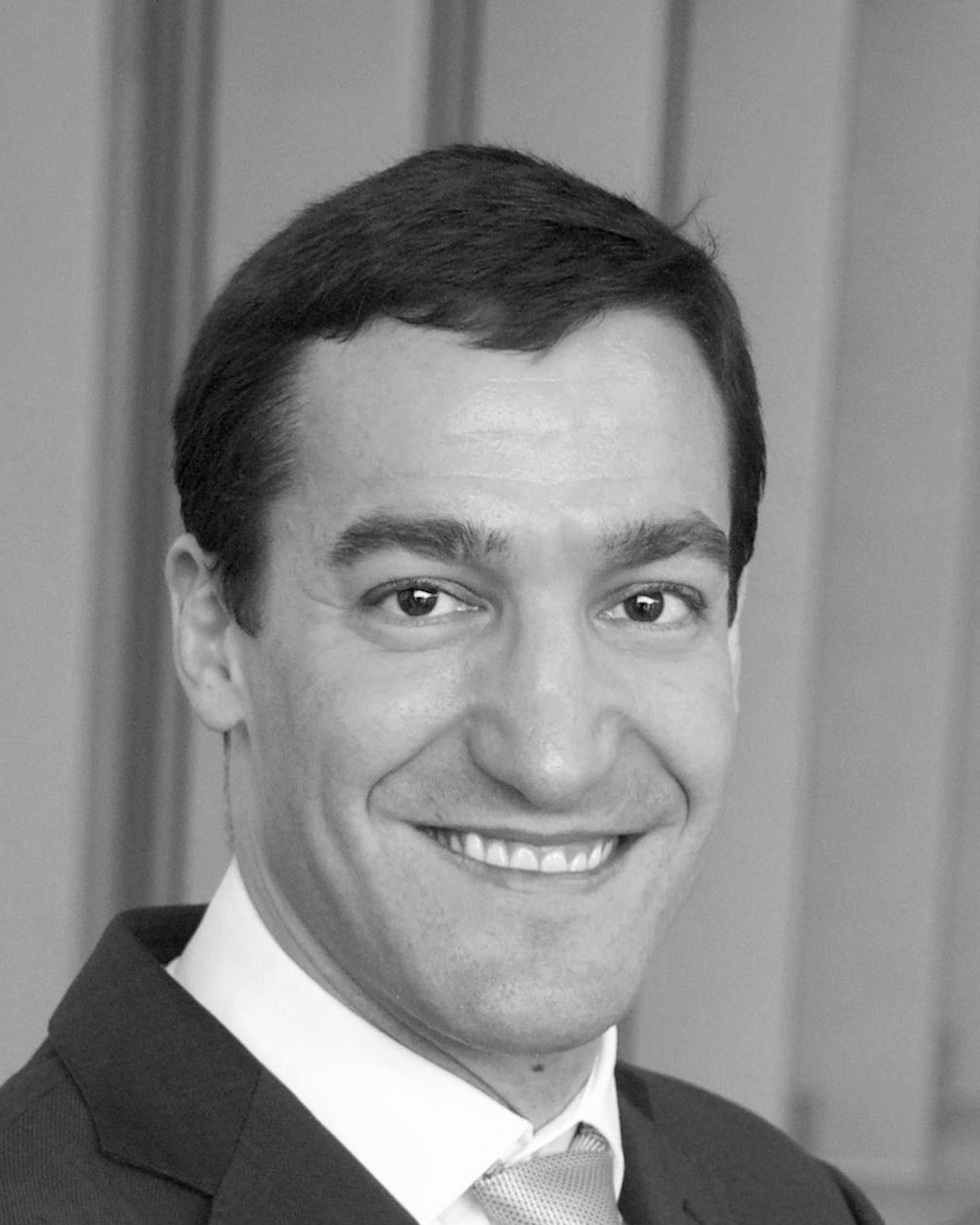}}]{Dr. David Atienza}
received the PhD degree in computer science and engineering from UCM, Spain, and IMEC, Belgium, in 2005.
He is a full professor of electrical and computer engineering, head of the Embedded Systems Laboratory (ESL) and the Associate Vice President for Research Centers and Technology Platforms at EPFL, Switzerland. His research interests include system-level design methodologies for multi-processor system-on-chip (MPSoC) servers, edge AI architectures, and low-power design of circuits and systems. He has co-authored more than 450 papers, one book, and 13 patents. Dr. Atienza has received, among other recognitions, the 2024 Test-of-Time Best Paper Award at the International Conference on Hardware/Software Codesign and System Synthesis (CODES+ISSS) for the most influential paper in the last 15 years, the ICCAD 10-Year Retrospective Most Influential Paper Award in 2020, the Most Influential DAC Under-40 Innovators Award in 2018, and an ERC Consolidator Grant in 2016. He is an ACM Fellow.
\end{IEEEbiography}

\vfill
\end{document}